\documentclass[sigconf, nonacm, pdfa]{acmart}

\newcommand\vldbdoi{10.14778/3836663.3836706}
\newcommand\vldbpages{3525-3537}
\newcommand\vldbvolume{19}
\newcommand\vldbissue{11}
\newcommand\vldbyear{2026}
\newcommand\vldbauthors{\authors}
\newcommand\vldbtitle{\shorttitle}
\newcommand\vldbavailabilityurl{https://github.com/HPI-Information-Systems/wf-optimization}
\newcommand\vldbpagestyle{empty}

\usepackage{algorithm}
\usepackage{algpseudocode}
\usepackage{multirow}
\usepackage{soul}
\usepackage[notext,nomath]{stix}

\usepackage{enumitem}
\usepackage{subcaption}
\usepackage{alltt}
\usepackage{tabularx}
\usepackage{hyperref}
\usepackage{mathtools}
\usepackage{xspace}
\usepackage{layouts}
\usepackage{datenumber}
\usepackage{multicol}

\usepackage{euflag}

\usepackage{siunitx}
\usepackage[capitalise,nameinlink,noabbrev]{cleveref}

\usepackage[a-2b]{pdfx}

\usepackage[normalem]{ulem}
\usepackage{tikz}
\usepackage{amsmath,amssymb}
\usetikzlibrary{arrows.meta,positioning,calc}
\usepackage{alltt}

\def\ojoin{\setbox0=\hbox{$\Join$}%
\rule[0.05ex]{.2em}{.4pt}\llap{\rule[1.05ex]{.2em}{.4pt}}}
\def\leftouterjoin{\mathbin{\ojoin\mkern-6.7mu\Join}}
\def\rightouterjoin{\mathbin{\Join\mkern-6.7mu\ojoin}}
\def\fullouterjoin{\mathbin{\ojoin\mkern-6.7mu\Join\mkern-6.7mu\ojoin}}

\newcommand{\eg}{e.\,g.,\ }
\newcommand{\ie}{i.\,e.,\ }
\newcommand{\WF}{\textsc{wf}\xspace}
\renewcommand{\WF}{\textsf{WF}\xspace}
\newcommand{\WFPLUS}{\textsc{wf+}\xspace}
\renewcommand{\WFPLUS}{\textsf{WF+}\xspace}
\renewcommand{\WFPLUS}{\textsf{WF\ensuremath{^+}}\xspace}
\newcommand{\WFs}{\textsc{wf}{\small s}\xspace}
\renewcommand{\WFs}{\textsf{WF}s\xspace}
\newcommand{\myparagraph}[1]{\smallskip\noindent\textbf{#1.}}
\newcommand{\punctfootnote}[1]{\unskip\nolinebreak\hspace{-.2em}\footnote{#1}}
\newcommand*{\landO}{\ensuremath{\text{\usefont{OMS}{cmsy}{m}{n}O}}}

\newcounter{equivalency}
\newcounter{proof}
\makeatletter
\newenvironment{equivalenc}[1]{%
\refstepcounter{equivalency}%
\label{equiv:#1}%
}{}

\makeatother

\usepackage{refcount}

\makeatletter
\newcommand{\setprimelabel}[2]{%
  \edef\crt@tmp{\getrefnumber{#1}}%
  \protected@edef\@currentlabel{\crt@tmp\ensuremath{'}}%
  \def\cref@currentlabel{[exquery][1][]\crt@tmp\ensuremath{'}}%
  \phantomsection
  \label{#2}%
}
\makeatother

\newcommand{\eq}[1]{\begin{equivalenc}{#1}(\arabic{equivalency})\end{equivalenc}}

\Crefname{equivalency}{Equivalence}{Equivalences}
\crefname{equivalency}{Equivalence}{Equivalences}
\Crefname{proof}{Proof}{Proofs}
\Crefname{proof}{Proof}{Proofs}

\newcommand{\opTheta}{\ensuremath{\mathbin{\theta}}}

\newcounter{exquery}
\renewcommand{\theexquery}{Q\arabic{exquery}}
\crefname{exquery}{}{} 
\Crefname{exquery}{}{}

\definecolor{colorA}{HTML}{88CCEE}
\definecolor{colorB}{HTML}{CC6677}
\definecolor{colorC}{HTML}{DDCC77}
\definecolor{colorD}{HTML}{117733}
\definecolor{colorE}{HTML}{332288}
\definecolor{colorF}{HTML}{AA4499}
\definecolor{colorG}{HTML}{44AA99}
\definecolor{colorH}{HTML}{999933}
\definecolor{colorI}{HTML}{882255}
\definecolor{colorJ}{HTML}{661100}
\definecolor{colorK}{HTML}{6699CC}

\newcommand{\revisionA}[2]{#2}
\newcommand{\revisionB}[2]{#2}
\newcommand{\revisionC}[2]{#2}
\newcommand{\revisionD}[2]{#2}
\newcommand{\revisionM}[2]{#2}
\newcommand{\revision}[1]{#1}

\newcommand{\revisionstyleA}[1]{#1}
\newcommand{\revisionstyleB}[1]{#1}

\newcommand{\revisionstyle}[2][]{#2}

\begin{document}

\title[Window Function Optimization: Co-Evaluation and Other Techniques]{Window Function Optimization:\\Co-Evaluation and Other Techniques}

\author{Daniel Lindner}
\orcid{0009-0003-1849-7262}
\affiliation{%
  \institution{Hasso Plattner Institute}
  \city{Potsdam, Germany}
}
\email{daniel.lindner@hpi.de}

\author{Felix Naumann}
\orcid{0000-0002-4483-1389}
\affiliation{%
  \institution{Hasso Plattner Institute}
  \city{Potsdam, Germany}
}
\email{felix.naumann@hpi.de}

\author{Alberto Lerner}
\orcid{0000-0003-4252-0648}
\affiliation{%
  \institution{Computing Flows}
  \city{Fribourg, Switzerland}
}
\email{lernera@computingflows.com}

\begin{abstract}
Window functions are among the most expressive features of modern SQL\@.
Surprisingly, relatively little has been written about their optimization.
Some techniques exist, such as pushing predicates through a window under ideal conditions, but known optimizations no longer apply when those conditions are even slightly unmet.
We show that these limitations are not fundamental, but persist because a reasoning framework for window function optimization has been missing.
We provide such a framework, introducing techniques we call Frame Analysis, Partition Analysis, and a new execution strategy called Co-Evaluation.
These clarify when and how optimizations can be applied.
Co-Evaluation, in particular, allows early evaluation of predicates even when they depend on the window function's result.
We present each technique and organize the results as a table of algebraic equivalences for window functions.
\revision{We test these optimizations in an open-source engine, where they never hurt performance and make certain common queries up to 40.7$\times$ faster, with larger tables yielding larger gains.}
\end{abstract}

\maketitle

\pagestyle{\vldbpagestyle}
\begingroup\small\noindent\raggedright\textbf{PVLDB Reference Format:}\\
    \vldbauthors. \vldbtitle. PVLDB, \vldbvolume(\vldbissue): \vldbpages, \vldbyear.
\href{https://doi.org/\vldbdoi}{doi:\vldbdoi}
\endgroup
\begingroup
\renewcommand\thefootnote{}\footnote{\noindent
This work is licensed under the Creative Commons BY-NC-ND 4.0 International License. Visit \url{https://creativecommons.org/licenses/by-nc-nd/4.0/} to view a copy of this license. For any use beyond those covered by this license, obtain permission by emailing \href{mailto:info@vldb.org}{info@vldb.org}. Copyright is held by the owner/author(s). Publication rights licensed to the VLDB Endowment. \\
\raggedright Proceedings of the VLDB Endowment, Vol. \vldbvolume, No. \vldbissue\ %
ISSN 2150-8097. \\
\href{https://doi.org/\vldbdoi}{doi:\vldbdoi} \\
}\addtocounter{footnote}{-1}\endgroup

\ifdefempty{\vldbavailabilityurl}{}{
\vspace{.3cm}
\begingroup\small\noindent\raggedright\textbf{PVLDB Artifact Availability:}\\
The source code, data, and/or other artifacts have been made available at \url{\vldbavailabilityurl}.
\endgroup
}

\section{Introduction}
\label{sec:intro}
Window functions (\WFs) provide explicit support for order-depen\-dent computations to be expressed in SQL~\cite{ISO:9075:2023} 
\revisionD{Meta}{and are commonly present in data analytics queries.
For instance, they appear in \qty{11}{\percent} of Snowfla\-ke's customer queries~\cite{DBLP:journals/pvldb/SzlangBCDFHOOM25}, and in Meta's reporting workloads, ``most query shapes contain joins, aggregations, or window functions''~\cite{DBLP:conf/icde/SethiTSPXSYJHSB19}.
In a case study we conducted on traces from two large SAP customers, \WF queries accounted for as much as \qty{9.5}{\percent} of total execution time.
}%
Nearly every modern database engine implements \revision{\WFs}~\cite{Oracle:WindowFunctions,MSSQL:WindowFunctions,DB2:WindowFunctions,PostgreSQL:WindowFunctions,MySQL:WindowFunctions,DuckDB:WindowFunctions,SQLite:WindowFunctions,Snowflake:WindowFunctions,ClickHouse:WindowFunctions,CockroachDB:WindowFunctions,AWS:Redshift:WindowFunctions,Google:BigQuery:WindowFunctions,CedarDB:SQLQueries,SingleStore:WindowFunctions,tidb:windowfunctions,oceanbase:windowfunctions,polardbx:windowfunctions}.

\revision{For this reason}, \WFs have been studied in the context of query optimization.
Cao et al. report on how to share partitioning and sorting work across multiple windows~\cite{DBLP:journals/pvldb/CaoCLT12}.
Leis et al. present a highly optimized parallel physical window operator for main-memory systems~\cite{DBLP:journals/pvldb/LeisKK015}.
More recently, \citeauthor{DBLP:journals/pvldb/Baca24} shows that rewriting certain windows as self-joins can be beneficial~\cite{DBLP:journals/pvldb/Baca24}.

These works have clearly advanced the field, but some gaps still remain. 
No table of algebraic equivalences involving \WFs exists, nor has a formalization been introduced that is precise enough to derive one. 
We provide both. 
Our formalization breaks \WFs into basic phases instead of capturing their entire semantics at once. 
Each phase is a precise yet simple transformation that builds on the output of its predecessor (\cref{sec:semantics}). 
This decomposition enables reasoning about intermediate execution states in a way that a monolith-style formalization could not offer. 
More importantly, it gives us insights into a number of details that are useful for algebraic optimizations
(\cref{sec:equivalences}).

Some of these equivalences are already used in practice for certain queries, but even slight query changes can render them inapplicable.
The challenge is not always to find new transformations, but to recognize when existing ones still apply.
We address this challenge through a series of techniques that extend the reach of basic transformations.
They comprise two \textit{analyses} that identify optimization opportunities and a new evaluation strategy: \textit{Co-Evaluation}.

The first technique is called \textit{Frame Analysis}.
It detects the conditions under which a \WF can be replaced by a simpler formulation~(\cref{sec:frame}).
For instance, a frame spanning a single row, as in the following query, does not need to evaluate the \WF.

\medskip
\begin{center}
\begin{tikzpicture}[
    every node/.style={anchor=west, inner sep=0pt, font=\ttfamily\small},
    node distance=0pt and 0pt
]
\node[font=\bfseries\footnotesize] (label) {\refstepcounter{exquery}\theexquery:\label{lst:q-frame}};
\node[right=0.2cm of label.east, anchor=west] (l1) {SELECT *, SUM(a) OVER (};
\node[below=0.8\baselineskip of l1.west, anchor=west] (l2) {\hspace{2em}ROWS BETWEEN CURRENT ROW AND CURRENT ROW)};
\node[below=0.8\baselineskip of l2.west, anchor=west] (l3) {FROM r};
\end{tikzpicture}
\end{center}
\medskip

Here, the \WF is \texttt{SUM(a) OVER(...)} and it specifies the frame, \ie the set of rows the aggregation considers, as \texttt{ROWS BETWEEN ... ROW}.
This query applies \texttt{SUM(a)} to each row individually, which is the value of attribute \texttt{a}.
A human would not write such a query, but software-written queries can feature such anomalies.
Without the \WF, this query runs 4$\times$ to 40.7$\times$ faster, depending on the table size.
Frame Analysis recognizes and replaces such singleton frames.

The second technique, \textit{Partition Analysis}, determines when predicates or joins can be pushed through a \WF~(\cref{sec:partition}).
Some of the techniques here are not new; predicates that match the \WF's partition expression exactly can benefit from early execution.
However, there are very common queries for which this alignment is not perfect, as in the following query.
\medskip
\begin{center}
\begin{tikzpicture}[
    every node/.style={anchor=west, inner sep=0pt, font=\ttfamily\small},
    node distance=0pt and 0pt
]
\node[font=\bfseries\footnotesize] (label) {\refstepcounter{exquery}\theexquery:\label{lst:q-partition}};
\node[right=0.2cm of label.east, anchor=west] (l1) {SELECT * FROM (};
\node[below=0.8\baselineskip of l1.west, anchor=west] (l2) {\hspace{2em}SELECT *, AVG(price) OVER (};
\node[below=0.8\baselineskip of l2.west, anchor=west] (l3) {\hspace{4em}PARTITION BY EXTRACT(YEAR FROM sold\_date)};
\node[below=0.8\baselineskip of l3.west, anchor=west] (l4) {\hspace{2em}) AS yearly\_avg};
\node[below=0.8\baselineskip of l4.west, anchor=west] (l5) {\hspace{2em}FROM sales) AS s};
\node[below=0.8\baselineskip of l5.west, anchor=west] (l6) {WHERE sold\_date >= DATE '2020-02-15'};
\end{tikzpicture}
\end{center}
\medskip

The query computes yearly average prices for recent sales.
It is written as an outer filtering query over an inner window query because the inner comes from a view while the outer is user-written.
Both the predicate and the window operate on \texttt{sold\_date}, so one may consider swapping them.
Filtering early would attenuate the heavy sorting on the window.
This predicate, however, alters the year 2020 partition, making the aggregation produce wrong results.

On close inspection, the predicate implies that only the partitions with a year $\geq 2020$ are relevant.
A predicate that filtered full years prior to 2020 could be correctly pushed.
Partition Analysis offers techniques to recognize this scenario and determine such \textit{derived predicates} that align with the partition expression.
On a historical sales database, for instance, this optimization can eliminate decades of partitions in a single pass.

In the example above, we have considered predicates that are independent of the window's output.
A harder problem arises when a predicate depends on it, as is the case for the following query~\ref{lst:q-coeval}.

\medskip
\begin{center}
\begin{tikzpicture}[
    every node/.style={anchor=west, inner sep=0pt, font=\ttfamily\small},
    node distance=0pt and 0pt
]
\node[font=\bfseries\footnotesize] (label) {\refstepcounter{exquery}\theexquery:\label{lst:q-coeval}};
\node[right=0.2cm of label.east, anchor=west] (l1) {SELECT * FROM (};
\node[below=0.8\baselineskip of l1.west, anchor=west] (l2) {\hspace{2em}SELECT *, RANK() OVER (};
\node[below=0.8\baselineskip of l2.west, anchor=west] (l3) {\hspace{4em}PARTITION BY dept};
\node[below=0.8\baselineskip of l3.west, anchor=west] (l4) {\hspace{4em}ORDER BY salary DESC) AS rnk};
\node[below=0.8\baselineskip of l4.west, anchor=west] (l5) {\hspace{2em}FROM employees) AS s};
\node[below=0.8\baselineskip of l5.west, anchor=west] (l6) {WHERE rnk <= 3};
\end{tikzpicture}
\end{center}
\medskip

The query filters out all but the top three earners per department.
The predicate \texttt{rnk <= 3} cannot be pushed through the \WF because the rank of each \texttt{salary} must first be calculated within a department.
However, we could discard salaries that cannot possibly rank among the top three early in the computation, substantially reducing the data the window processes.
We call this technique \textit{Co-Evaluation}~(\cref{sec:coevaluation}), and in our experiments, \cref{lst:q-coeval}'s performance can run up to 5$\times$ faster with this optimization.

A standard \WF operator has no place to inject this early pruning.
A natural, relatively easy place to add it would be alongside the aggregation, but that yields little benefit.
By then, the data has already been sorted, and sorting dominates the cost.
The real gains come from pruning before or during sorting.
We show how a standard \WF operator can be extended to prune early; we call this \WFPLUS and describe our implementation in an open-source database~(\cref{sec:implementation}).

To measure the benefits, we evaluate our proposed techniques on queries that exemplify the optimization gaps we address. 
We ran them on tables with varying sizes, partition count, and skew to reveal which factors influence performance~(\cref{sec:experiments}). 
As mentioned, some experienced speedups between one and two orders of magnitude.
We close the paper by comparing our techniques to related work~(\cref{sec:related}) and by stating our final remarks~(\cref{sec:conclusion}).
In summary, the main contributions of this paper are:
\begin{itemize}[leftmargin=*]
  \item A formal algebraic definition of window functions.
  \item A formal table of algebraic equivalences involving \WFs.
  \item Frame and Partition Analyses that enable new optimizations or apply known optimizations in new scenarios.
  \item Implementation notes on how to incorporate the equivalences and analyses \revision{into} existing engines.
  \item Co-Evaluation, a technique that fuses predicates with window execution to early prune non-qualifying rows.
  \item An experimental evaluation that applies the optimizations in practice and quantifies the speedups they bring.
\end{itemize}

\section{Window Function Semantics}
\label{sec:semantics}
This section formally defines the window function operator.
We first establish notation, then describe the operator as a sequence of phases, and finally show how SQL's syntax maps to this model.
\revisionA{R1.O1}{We define the operator over sequences rather than within relational algebra because relational algebra cannot represent several features required to capture the semantics of \WFs, most notably order.}

\subsection{Preliminaries}
\label{sec:prelim}

\begin{figure*}[t]
\centering
\resizebox{\textwidth}{!}{
\input{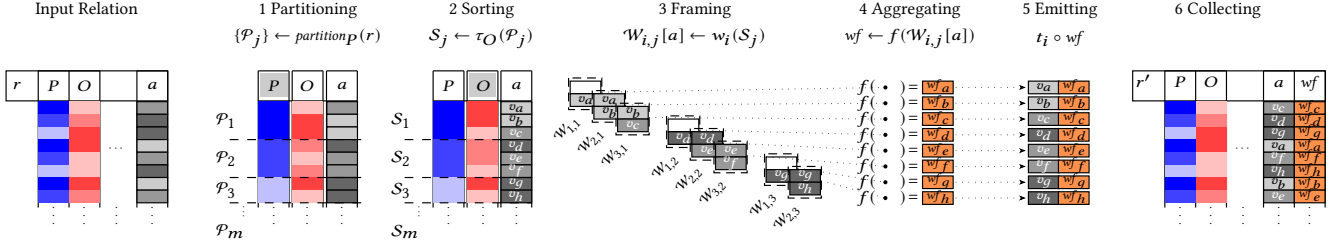}
}
\vspace{-2em}
\caption{Phases of a window function operator: from input (left) to outputting a \WF column (right).}
\vspace{-0.5em}
\label{fig:wf_semantics}
\end{figure*}

We use standard notation for relational operations and introduce symbols that are specific to window functions in \cref{tab:notation}.

\myparagraph{Relations and tuples}
A \emph{relation} $r$ is a bag (multiset) of tuples.
We write $\mathcal{A}(r)$ for the set of attributes of $r$.
For a tuple $t$ and attribute set $A \subseteq \mathcal{A}(r)$, $t[A]$ denotes the projection of $t$ onto $A$.
Two tuples \emph{agree} on $A$ if they have the same values for all attributes in $A$.
Tuple concatenation $t \circ u$ appends the attributes of $u$ to $t$.
We write $\theta$ for a generic comparison operator ($=$, $<$, $\leq$, $>$, $\geq$) and $a \opTheta c$ for a predicate comparing attribute $a$ to constant $c$.

\myparagraph{Sequences}
A \emph{sequence} $\mathcal{S} = \langle t_1, \ldots, t_n \rangle$ is an ordered list of tuples~\cite{DBLP:journals/jacm/AbiteboulG86}.
We write $|\mathcal{S}|$ for its length, $||\mathcal{S}||$ for its number of distinct elements, 
$\mathcal{S}[k]$ for the $k^{\text{th}}$ tuple (1-indexed), and $\mathcal{S}[l:u]$ for the subsequence from $l$ to $u$, inclusive.
$\mathcal{S}[l:u] = \langle \rangle$ if $l > u$.
For index sequences $\mathcal{I} = \langle k_1, k_2, \ldots \rangle$, $\mathcal{S} - \mathcal{I}$ removes the positions in $\mathcal{I}$ from $\mathcal{S}$.
We write $\langle k : C(k) \rangle$ for the sequence of positions $k$ satisfying condition $C$.
$\mathcal{S}[\mathcal{I}]$ denotes the sequence $\langle t_{k_1}, t_{k_2}, \ldots \rangle$. 
$\mathcal{S}[\mathcal{I}][A]$ is the projection $\langle t_{k_1}[A], t_{k_2}[A], \ldots \rangle$.
Concatenation is written $\mathcal{S}_1 \circ \mathcal{S}_2$.

\myparagraph{Dependencies}
An attribute set $X$ is a \emph{unique column combination} (UCC)  if no tuple pair agrees on $X$, \ie $X$ uniquely identifies each tuple.
A \emph{functional dependency} (FD) $X \rightarrow Y$ holds on relation $r$ if for all tuple pairs $t_1, t_2 \in r$: $t_1[X] = t_2[X] \Rightarrow t_1[Y] = t_2[Y]$.
An \emph{order dependency} (OD) $[X] \mapsto [Y]$ holds if ordering by $X$ implies that $Y$ is ordered~\cite{DBLP:journals/pvldb/SzlichtaGGZ13}, 
\eg $[\textit{date}] \mapsto [\textit{year}]$.

\myparagraph{Relational operators}
We use standard notation.
Selection $\sigma_C(r) = \{t \in r \mid C(t)\}$ returns all tuples of $r$ satisfying predicate $C$~\cite[p.~56]{DBLP:books/cs/Ullman88}.
Projection $\pi_E(r)$ computes expressions $E$ for each tuple, not limited to attribute names~\cite[p.~217f]{DBLP:books/daglib/0020812}.
Sort $\tau_O(r)$ orders tuples by attribute list $O$~\cite[p.~219]{DBLP:books/daglib/0020812}.
Limit $\lambda_k(r)$ returns the first $k$ tuples~\cite{DBLP:conf/sigmod/LiCIS05}.
Group-by $\gamma_{G,f}(r)$ partitions $r$ by attributes $G$ and computes aggregate $f$ over each group, returning one tuple per group.

For join variations, we adopt~\cite[pp.~45--58, 220f]{DBLP:books/daglib/0020812} as follows:
inner join is $r \Join_C s = \sigma_C(r \times s) = \{ t \circ u \mid t \in r \land u \in s \land C(t \circ u) \}$.
Semi-join is $r \ltimes_C s = \{t \in r \mid \exists u \in s: C(t \circ u)\}$, returning tuples from $r$ that have at least one match in $s$.
Anti-join is $r \triangleright_C s = \{t \in r \mid \nexists u \in s: C(t \circ u)\}$, returning tuples from $r$ with no match in $s$; tuples with \textsf{NULL} values in the join key are preserved.
Left outer join is $r \leftouterjoin_C s = (r \Join_C s) \cup \{t \circ [\bot, \ldots] \mid t \in r \triangleright_C s\}$, which preserves all tuples from $r$, padding with \textsf{NULL}s when no match exists.
Right outer ($\rightouterjoin$) and full outer ($\fullouterjoin$) joins are defined analogously.

\myparagraph{Comparison semantics}
Lexicographical comparisons $\preccurlyeq$ in the context of partitioning, grouping, and frame peer determination follow SQL's \emph{syntactic equality} semantics, under which \textsf{NULL} values are treated as equal to each other~\cite{DBLP:conf/pods/LibkinP23,ISO:9075:2023}.
Thus, \textsf{NULL}~$=$~\textsf{NULL}, \textsf{NULL}~$\leq$~\textsf{NULL}, and \textsf{NULL}~$\geq$~\textsf{NULL} all evaluate to true in these contexts, matching the
SQL standard.
Furthermore, it can be specified whether \textsf{NULL}s are sorted first or last in the result.

\subsection{The Window Function Operator}
\label{ssec:formal}

We denote the window function operator by $\boxplus_{P,O,w,f}(r)$, where $P$ is a set of partition attributes, $O$ is a list of order attributes, $w$ is a frame function, and $f$ is an aggregate function.\punctfootnote{The aggregate operates on attribute $a$, which we keep implicit in the same way that aggregation $\gamma_{G,f}(r)$ does. The $\boxplus$ symbol is produced by \texttt{\textbackslash boxplus} in \LaTeX.}
The window function operator is defined as:
\begin{align*}
\boxplus_{P,O,w,f}(r) \;\;&\coloneq\;\; \bigcup_{j=1}^{m} \Big\{ t_i \circ f\big(\mathcal{W}_{i,j}[a]\big) \;\Big|\; i \in \{1, \ldots, |\mathcal{S}_j|\} \Big\} \\[0.5em]
\makebox[0pt][r]{\text{where}\hspace{2em}} \;\;&\\[0.5em]
\mathcal{W}_{i,j} \;\;&\coloneq\;\; \mathcal{S}_j[w_i(\mathcal{S}_j)] \tag*{(framing)\hspace{0.5cm}}\\
\mathcal{S}_j \;\;&\coloneq\;\; \tau_O(\mathcal{P}_j) \tag*{(sorting)\hspace{0.5cm}}\\
\{\mathcal{P}_1, \ldots, \mathcal{P}_m\} \;\;&\coloneq\;\; \textit{partition}_P(r) \tag*{(partitioning)\hspace{0.5cm}}
\end{align*}\smallskip

Informally, the operator transforms relation $r$ through a sequence of phases, as depicted by~\cref{fig:wf_semantics}.
First, tuples are grouped into $m$ partitions $\mathcal{P}_1, \ldots, \mathcal{P}_m$ based on $P$.
Each partition $\mathcal{P}_j$ is sorted by $O$, producing an ordered sequence $\mathcal{S}_j$.
For each tuple $t_i$ in sequence $\mathcal{S}_j$, a frame function $w$ builds the window frame for that tuple, $\mathcal{W}_{i,j}$, and an aggregate function $f$ computes the window function result, $\mathit{wf}$, over that frame.
As a result, the original tuple is extended with the computed value.
Finally, results from all partitions are collected back into a (bag) relation.
We now define each phase formally.

\begin{table}[t]
\centering
\caption{Selected notation for window function components.}
\vspace{-1em}
\label{tab:notation}
\small\begin{tabular}{@{}ll@{}}
\toprule
\textbf{Symbol} & \textbf{Description} \\
\midrule
$X \rightarrow Y$ & Functional dependency over attribute lists $X$ and $Y$\\
$[X] \mapsto [Y]$ & Order dependency: ordering by $X$ implies ordering by $Y$ \\
\midrule
$\mathcal{S}$ & Sequence of tuples \\
$|\mathcal{S}|$ & Length of sequence $\mathcal{S}$ \\
$||\mathcal{S}||$ & Number of distinct elements in $\mathcal{S}$ \\
$\mathcal{S}[k]$ & Element at position $k$ (1-indexed) \\
$\mathcal{S}[l:u]$ & Subsequence from position $l$ to $u$ (inclusive) \\
$\langle k_1, \ldots, k_n \rangle$ & Elements of index sequence $\mathcal{I}$ \\
$\langle k : C(k) \rangle$ & Sequence of indexes $k$ satisfying condition $C$ \\
$\mathcal{S}[\mathcal{I}]$ & Subsequence at positions in index sequence $\mathcal{I}$ \\
$\mathcal{S} - \mathcal{I}$ & Sequence $\mathcal{S}$ with positions in $\mathcal{I}$ removed \\
$\mathcal{S}_1 \circ \mathcal{S}_2$ & Concatenation of sequences \\
\midrule
$\boxplus_{P,O,w,f}(r)$ & Window function operator \\
$P$ & Partition attributes \\
$O$ & Order attributes (with direction, \textsf{NULL} ordering) \\
$w$ & Frame function \\
$f$ & Aggregate or ranking function over implicit attribute $a$ \\
$\{\mathcal{P}_1, \ldots, \mathcal{P}_m\}$ & Set of partitions of a relation \\
$\mathcal{S}_j$ & Sorted sequence for partition $\mathcal{P}_j$ \\
$\mathcal{W}_{i,j}$ & Frame contents for tuple $i$ in partition $j$ \\
\bottomrule
\vspace{-1.5em}
\end{tabular}
\end{table}

\myparagraph{Phase 1: Partitioning}
$\textit{partition}_P(r) = \{\mathcal{P}_1, \ldots, \mathcal{P}_m\}$, where $m$ is the number of distinct values of $P$ in $r$, $\mathcal{P}_j \cap \mathcal{P}_k = \varnothing$ for $j \neq k$, every tuple $t \in r$ belongs to exactly one $\mathcal{P}_j$, and $\forall t, t' \in \mathcal{P}_j\colon t[P] = t'[P]$.
The standard treats $P$ as a list of attributes, but in practice, engines allow expressions over these attributes.
We assume the latter.

\myparagraph{Phase 2: Sorting}
For each $\mathcal{P}_j$, $\tau_O(\mathcal{P}_j) = \mathcal{S}_j = \langle t_1, \ldots, t_n \rangle$, where $\{t_1, \ldots, t_n\} = \mathcal{P}_j$ and $\forall i \in \{1, \ldots, n{-}1\}\colon t_i[O] \preccurlyeq  t_{i+1}[O]$.
When $O$ does not uniquely determine order, multiple valid sequences exist; all equivalences in this paper hold for every valid ordering.

\myparagraph{Phase 3: Framing}
For each position $i$ in $\mathcal{S}_j$, $w_i(\mathcal{S}_j)$ is a sequence of indexes $\langle k_1, k_2, \ldots\rangle$ into $\mathcal{S}_j$ that reduces the partition to the elements that should be considered for aggregation.
The frame contents for $t_i$ are $\mathcal{W}_{i,j} = \mathcal{S}_j[w_i(\mathcal{S}_j)] = \langle t_{k_1}, t_{k_2}, \ldots \rangle$.
If an exclusion modifier is specified, it filters the positions accordingly.

\myparagraph{Phase 4: Aggregating}
$f\colon \langle t_{k_1}, t_{k_2}, \ldots \rangle \rightarrow \mathit{wf}$, a function that computes a value $\mathit{wf}$ from $\mathcal{W}_{i,j}[a]$.

\myparagraph{Phase 5: Emitting}
For each tuple $t_i$ in $\mathcal{S}_j$, emit $t_i \circ f(\mathcal{W}_{i,j}[a])$, the original tuple extended with the computed value.

\myparagraph{Phase 6: Collecting}
The result is $\bigcup_{j=1}^{m} \{ t_i \circ f(\mathcal{W}_{i,j}[a]) \mid i \in \{1, \ldots, |\mathcal{S}_j|\} \}$.
The ordering imposed by phase 2 is not preserved; the output is a bag with no ordering guarantees.

\subsection{From SQL to the Window Function Operator}
\label{ssec:sql_to_wf}

SQL's \texttt{OVER()} clause~\cite{ISO:9075:2023} translates directly into the operator's four parameters:
\begin{center}
\begin{tabular}{@{}l@{\quad}l@{}}
\textit{f}\texttt{(a) OVER (} & $\rightarrow f, a$\\
\quad\texttt{PARTITION BY ...} & $\rightarrow P$ \\
\quad\texttt{ORDER BY ...} & $\rightarrow O$ \\
\quad\textit{frame\_clause} & $\rightarrow w$ \\
\texttt{)} & \\
\end{tabular}
\end{center}

Obtaining $P$, $f$, and $a$ from SQL is straightforward.
The order list $O$, however, encodes not just attributes but also sort direction and \textsf{NULL} ordering (\texttt{ASC}/\texttt{DESC}, \texttt{NULLS FIRST}/\texttt{LAST}).
The frame function $w$ captures the richest semantics: frame mode (\texttt{ROWS}, \texttt{RANGE}, \texttt{GROUPS}), boundaries, and exclusion.
Translating SQL's various frame clauses into our sequence notation requires care.
\cref{tab:frames} illustrates the mapping \revisionB{R2.W4, R2.D4}{that generalizes to all} cases.
\revisionstyleB{Note that the SQL standard allows only a single order attribute for \texttt{RANGE} mode.}

\begin{table}[ht]
\centering
\caption{Frame clauses to frame functions examples.}
\label{tab:frames}
\begin{tabularx}{\columnwidth}{@{}X@{}}
\toprule
\textbf{Frame clause} $\rightarrow w_i(\mathcal{S})$ \\
\midrule
\texttt{ROWS BETWEEN UNBOUNDED PRECEDING AND CURRENT ROW} \\
\hfill $\rightarrow \langle 1, \ldots, i \rangle$ \\[1ex]
\texttt{ROWS BETWEEN $n$ PRECEDING AND $m$ FOLLOWING} \\
\hfill $\rightarrow \langle \max(1, i-n), \ldots, \min(|\mathcal{S}|, i+m) \rangle$ \\[1ex]
\texttt{GROUPS BETWEEN UNBOUNDED PRECEDING AND CURRENT ROW} \\
\hfill $\rightarrow \langle k: k \leq i \lor \mathcal{S}[k][O] = \mathcal{S}[i][O] \rangle$ \\[1ex]
\texttt{GROUPS BETWEEN $n$ PRECEDING AND $m$ FOLLOWING} \\
\hfill $\rightarrow \langle k: 0 < ||\mathcal{S}[k:i][O]|| \leq n + 1 \lor 0 < ||\mathcal{S}[i:k][O]|| \leq m + 1  \rangle$ \\[1ex]
\texttt{RANGE BETWEEN UNBOUNDED PRECEDING AND CURRENT ROW} \\
\hfill $\rightarrow \langle k : \mathcal{S}[k][O] \leq \mathcal{S}[i][O] \rangle$ \\[1ex]
\texttt{RANGE BETWEEN $u$ PRECEDING AND $v$ FOLLOWING} \\
\hfill $\rightarrow \langle k : 0 \leq\mathcal{S}[i][O] - \mathcal{S}[k][O] \leq u \lor 0 \leq \mathcal{S}[k][O] - \mathcal{S}[i][O] \leq v \rangle$ \\
\midrule
\textbf{Exclusion} $\rightarrow$ \textbf{filter on} $w$ \\
\midrule
\texttt{EXCLUDE CURRENT ROW} \hfill $\rightarrow w_i(\mathcal{S}) - \langle i \rangle$ \\[1ex]
\texttt{EXCLUDE GROUP} \hfill $\rightarrow w_i(\mathcal{S}) - \langle k : \mathcal{S}[k][O] = \mathcal{S}[i][O] \rangle$ \\[1ex]
\texttt{EXCLUDE TIES} \hfill $\rightarrow w_i(\mathcal{S}) - \langle k : k \neq i \land \mathcal{S}[k][O] = \mathcal{S}[i][O] \rangle$ \\
\bottomrule
\end{tabularx}
\end{table}

\section{Algebraic Equivalences}
\label{sec:equivalences}
Now that we have established a formal definition of the \WF operator, we can begin developing algebraic equivalences for it.
The idea is familiar from other operators: transform a relational expression into one that (hopefully) executes faster while producing the same results.
We make no claim to completeness but believe that the equivalences we present cover enough diverse cases to show applicability, and that the examples are representative.
We hope this framework will invite further research.

\begin{table*}[ht]
\centering
\caption{Algebraic equivalences for window function operators.}
\vspace{-1em}
\label{tab:transformations}
\begin{tabular}{p{2cm}r@{~}lp{8.65cm}}
\toprule
\textbf{Class} & \multicolumn{2}{l}{\textbf{Subclass/Equivalences}} & \textbf{Preconditions and Notes}
\\
\midrule
 \multirow{5}{=}{Attribute reductions (\cref{ssec:reduction})}
    & \multicolumn{2}{l}{Reduce partition by attributes} \\
    & \eq{reduce-partition}
    & $\boxplus_{P, \ldots}(r) \equiv \boxplus_{P \setminus \{p\}, \ldots}$
        & FD $P \setminus \{p\} \rightarrow p$  \\
    \cmidrule{2-4}
    & \multicolumn{2}{l}{Reduce order by attributes}\\
    & \eq{reduce-order1}
    & $\boxplus_{P, O, \ldots}(r) \equiv \boxplus_{P, O[1:k-1] \circ O[k+1:n], \ldots}(r)$
    & FD $\{o_1, \ldots, o_{k-1}\} \rightarrow o_k$\\
    & \eq{reduce-order2}
    & $\boxplus_{P, O, \ldots}(r) \equiv \boxplus_{P, O[2:n], \ldots}(r)$
    & FD $P \rightarrow o_1$\\
\midrule
\multirow{5}{=}{Operator substitutions (\cref{ssec:substitutions,sec:frame})}
    & \multicolumn{2}{l}{Replace with group by} \\
    &\eq{replace-groupby1}& $\boxplus_{\varnothing, \ldots, f}(r) \equiv r \times \gamma_f(r)$ & \multirow{2}{=}{Equivalence per definition (\cref{sec:prelim}) if $f$ is standard aggregate function and frame covers entire partition.} \\
    &\eq{replace-groupby2}& $\boxplus_{P, \ldots, f}(r) \equiv r \Join_{P = P} \gamma_{P, f}(r)$  \\
    \cmidrule{2-4}

    & \multicolumn{2}{l}{Replace with projection \revisionstyleA{(Frame Analysis)}}\\
    & \eq{replace-projection}
    & $\boxplus_{\ldots}(r) \equiv \pi_{\mathcal{A}(r)\circ e}(r)$
    & All windows contain either only the current or no tuple. \\

\midrule
\multirow{8}{=}{Push-downs (\revisionstyleA{Partition Analysis,} \cref{sec:partition})}
    & \multicolumn{2}{l}{Predicate push-down}
    \\
    & \eq{pd-pred1}
    & $\sigma_{b \opTheta \textit{c}}(\boxplus_{P, \ldots}(r)) \equiv \boxplus_{P, \ldots}(\sigma_{b \opTheta \textit{c}}(r))$
    & FD $P \rightarrow b$ \\
    & \eq{pd-pred2}
    & $\sigma_{b \opTheta \textit{c}}(\boxplus_{P, \ldots}(r)) \equiv \sigma_{b \opTheta \textit{c}}(\boxplus_{P, \ldots}(\sigma_{p \opTheta \textit{c}'}(r)))$
    & FD $\{b\} \rightarrow p \in P$, $c' \in \pi_p(r), b \opTheta c \Rightarrow p \opTheta c'$
     \\
    \cmidrule{2-4}

    & \multicolumn{2}{l}{Join push-down} & $P, O, w, f$ reference only $\mathcal{A}(r)$ and:\\
    &\eq{pd-join1}
    & $\boxplus_{P,\ldots}(r) \Join_{r.b=s.x} s \equiv \boxplus_{P,\ldots}(r \Join_{r.b=s.x} s)$
    & FD $P \rightarrow b$ and $x$ unique \\
    &\eq{pd-join2} & $\boxplus_{P,\ldots}(r) \leftouterjoin_{r.b=s.x} s \equiv \boxplus_{P,\ldots}(r \leftouterjoin_{r.b=s.x} s)$ & $x$ unique \\
    &\eq{pd-join3}
    & $\boxplus_{P,\ldots}(r) \ltimes_{r.b=s.x} s \equiv \boxplus_{P,\ldots}(r \ltimes_{r.b=s.x} s)$ & FD $P \rightarrow b$ \\
    &\eq{pd-join4}
    & $\boxplus_{P,\ldots}(r) \triangleright_{r.b=s.x} s \equiv \boxplus_{P,\ldots}(r \triangleright_{r.b=s.x} s)$ & FD $P \rightarrow b$ \\

\midrule
\multirow{5}{=}{Limit-like optimizations (\cref{sec:coevaluation})}
    & \multicolumn{2}{l}{Predicate elimination} \\
    &\eq{result-limit}& $\sigma_{f \leq c}(\boxplus_{\varnothing,O,w,f}(r)) \equiv \lambda_c(\tau_O(\boxplus_{\varnothing,O,w,f}(r))) $
    & (i)~$f = \texttt{row\_number()}$ or (ii)~$f \in \{\texttt{rank()}, \texttt{dense\_rank()}\}$ and $O$ is UCC. For \revision{a} pure sort-based \WF operator, no extra sort $\tau_O$ required. 
    \\
    \cmidrule{2-4}
    & \multicolumn{2}{l}{Predicate co-evaluation} \\
    &\eq{co-eval}& $\sigma_{C}(\boxplus_{P, \ldots}(r)) \equiv \boxplus^+_{P, \ldots, C}(r)$
    & $C$ is any predicate; only MOD predicates benefit from early pruning.
    \\
\bottomrule
\end{tabular}
\end{table*}

\subsection{Equivalence Classes}
\label{ssec:classes}

We divide equivalences into four classes based on the techniques they use, as \Cref{tab:transformations} shows.
We summarize them here; later sections develop each in detail.
\begin{itemize}[leftmargin=*]

\item \textit{Reductions} target the partition and order clauses.
When functional dependencies hold, some attributes in these clauses contribute nothing, because they are constant within the scope where they would be compared.
Removing them shrinks the partition or sort keys, with no change to the result.

\item \textit{Substitutions} apply when the window falls into some special cases, \eg a frame has one or zero tuples or it takes the entire partition.
If the frame contains at most one tuple, the result depends only on the current row, and a projection suffices.
If the frame spans the entire partition, a \texttt{GROUP BY} and a join can replace the window.

\item \textit{Push-downs} swap selections or joins with the window.
The key observation here is that a window requires complete partitions to compute correctly.
An operator can be pushed only if it removes entire (non-relevant) partitions or leaves them intact.
We call this property partition integrity.
Some functional dependencies and uniqueness constraints can help determine when it holds.

\item \textit{Limit-like optimizations} address predicates that depend on the window's output.
Such predicates cannot be pushed, but some exhibit a useful structure: once false, they stay false for subsequent tuples in partition order.
This monotonicity implies a bound on rows that can qualify, enabling aggressive early pruning.
\end{itemize}

\revisionB{R2.D2}{The classes differ in how they relate to cost. 
Reductions (\cref{equiv:reduce-partition,equiv:reduce-order1,equiv:reduce-order2}), projection substitution (\cref{equiv:replace-projection}), full-relation substitution (\cref{equiv:replace-groupby1}), and limit elimination (\cref{equiv:result-limit}) never produce a slower plan, so an optimizer can apply them as unconditional rules. 
Predicate and reducing-join push-downs (\cref{equiv:pd-pred1,equiv:pd-pred2,equiv:pd-join3,equiv:pd-join4}) only shrink the window's input, so they are also safe to apply directly. 
An optimizer can apply these rules during normalization instead of treating them as alternatives that require enumeration and costing, thus avoiding any increase of the cost-based search space.
The remaining equivalences (\cref{equiv:replace-groupby2,equiv:pd-join1,equiv:pd-join2,equiv:co-eval}) are cost-dependent, and we identify their cost drivers where each is developed.}
We treat reductions and some substitutions in the next section, as they follow directly from the definitions in \cref{sec:semantics}.
The remaining equivalences require the analysis techniques we present in more detail in \cref{sec:frame,sec:partition,sec:coevaluation}.

\subsection{Reductions}
\label{ssec:reduction}

The partition and order clause lengths have some impact on the work done by a window operator.
Fewer partition attributes mean a narrower key to hash and compare.
The same applies to sorting.
FDs offer an opportunity here: when they make a partition or sort attribute redundant, we can remove it without affecting the result.

Consider \cref{equiv:reduce-partition}.
If the remaining partition attributes $P \setminus \{p\}$ functionally determine $p$, then tuples that agree on $P \setminus \{p\}$ must also agree on $p$.
The partition boundaries are identical with or without $p$, so any such attributes can be dropped.
This reduction affects only phase~1 in~\Cref{fig:wf_semantics}. 
As it delivers the same results for that phase, it can be safely applied to the \WF.

A similar line of reasoning supports \cref{equiv:reduce-order1,equiv:reduce-order2}.
Recall that sorting is lexicographic for compound attributes: attribute $o_k$ is compared only among tuples that already agree on $o_1, \ldots, o_{k-1}$.
If an FD makes $o_k$ constant within such groups, it contributes nothing to the comparison and can be removed (\cref{equiv:reduce-order1}).
Similarly, when the partition attributes determine the first ordering attribute, that attribute is constant within every partition and can be dropped (\cref{equiv:reduce-order2}).
These optimizations affect only phase~2 in~\Cref{fig:wf_semantics} without altering the results of the sort or affecting other phases.
As we hinted above, we expect this class to contain many more equivalences and dependency types than we show here.

\myparagraph{Implementation note}
Uncovering \emph{FD-based redundancies} can be done at optimization time.
For example, if attributes $a$ and $b$ are used together in a partitioning or sorting clause, but the query applies an equality filter to $b$ (turning it into a constant), the FD $\{a,b\} \setminus \{b\} \rightarrow b$ holds~\cite{DBLP:journals/vldb/KossmannPN22}.

\subsection{General Substitutions}
\label{ssec:substitutions}

Some windows reduce to trivial cases and can be replaced by simpler operators.
This can occur when the frame covers entire partitions, a case we address now, or when it covers one row or none, which we defer to Frame Analysis~(\cref{sec:frame}).

In \Cref{equiv:replace-groupby1}, the partition clause is missing, which is equivalent to eliminating phase~1 in~\Cref{fig:wf_semantics}.
More simply put, $m = 1$.
When $f$ is an aggregate function (as opposed to a ranking function), and the frame covers the entire partition, the \WF computes a single aggregate over the entire relation and attaches it to every row.
This is exactly what a Cartesian product with a scalar aggregation produces, including producing results for \textsf{NULL} values.

In \Cref{equiv:replace-groupby2}, a partition clause is present, so each partition produces its own aggregation result.
A similar optimization still applies, but the result of the aggregation must be joined back to the original tuples on the partition key.
Once again, the join uses syntactic equality (see \cref{sec:prelim}), matching SQL's semantics where \textsf{NULL} values are placed into the same partition.
This optimization is a bit nuanced, so we offer reasoning about its correctness next.

\myparagraph{Proof sketch for \cref{equiv:replace-groupby2}}
When $w_i(\mathcal{S}_j) = \langle 1, \ldots, |\mathcal{S}_j| \rangle$ for all $i$, the frame spans the entire partition.
By phase~4, aggregate~$f$ receives all tuples sharing the same $P$-value and produces a single value for each partition.
The aggregation $\gamma_{P,f}(r)$ computes the same aggregate per $P$-value and returns one tuple per group.
Joining $r$ with $\gamma_{P,f}(r)$ on $P$ replicates each aggregate to the original tuples.
For correctness, the join must treat \textsf{NULL}s in $P$ as equal, matching the partitioning semantics. \hfill$\square$

\revisionB{R2.D2}{Unlike reductions, this substitution is cost-dependent. The aggregation and the join on the partition key can cost more or less than the replaced window, depending on the number of partitions and their cardinalities. An engine should compare both versions.}

\section{Frame Analysis}
\label{sec:frame}
We begin with a motivating example.
Recall \cref{lst:q-frame} from the introduction.
The frame clause is \texttt{ROWS BETWEEN CURRENT ROW AND CURRENT ROW}, which contains exactly one row.
The window function \texttt{SUM(a)} applied to a single row simply returns \texttt{a}'s value.
Frame Analysis recognizes this and rewrites the query with \cref{equiv:replace-projection}:

\begin{center}
\begin{tikzpicture}[
    every node/.style={anchor=west, inner sep=0pt, font=\ttfamily\footnotesize},
    node distance=0pt and 0pt
]
\node[font=\bfseries\footnotesize] (label) {\ref{lst:q-frame}$^\prime$:\setprimelabel{lst:q-frame}{lst:q-frame-prime}};
\node[right=0.2cm of label.east, anchor=west] (l1a) {SELECT *,};
\node[right=0.3em of l1a.east, anchor=west, fill=gray!30, inner sep=2pt] (l1b) {a};
\node[right=0.3em of l1b.east, anchor=west] (l1c) {FROM r};
\end{tikzpicture}
\end{center}

To use this equivalence, we must determine (a)~how to recognize that a frame is singleton or empty, and (b)~what expression replaces the window function. 
The first case is clear from syntax, but others require careful Frame Analysis. 
We first address (a), as it is a precondition to applying \cref{equiv:replace-projection}. 
Then, we discuss (b), also needed in that equivalence.

\subsection{Detecting Singleton or Empty Frames}

A frame is guaranteed to contain only the current row under \revision{at least one of} four alternative conditions:
\begin{enumerate}[label=(\roman*),leftmargin=*]

\item Frame is \texttt{ROWS BETWEEN CURRENT ROW AND CURRENT ROW} or
\texttt{ROWS BETWEEN 0 PRECEDING AND 0 FOLLOWING}.
By \cref{tab:frames}, $w_i(\mathcal{S}_j) = \langle \max(1, i{-}0), \ldots, \min(|\mathcal{S}_j|, i{+}0) \rangle = \langle i \rangle$.

\item Frame is \texttt{RANGE/GROUPS BETWEEN CURRENT ROW AND CUR\-RENT ROW} with \texttt{EXCLUDE TIES}.
By Table~\ref{tab:frames}, $w_i(\mathcal{S}_j) = \langle k : \mathcal{S}_j[k][O] = \mathcal{S}_j[i][O] \rangle$, the set of peers.
\texttt{EXCLUDE TIES} removes all peers except the current row, leaving $\langle i \rangle$.

\item Frame is \texttt{RANGE/GROUPS BETWEEN CURRENT ROW AND CUR\-RENT ROW} and $O$ is a unique column combination.
The peer sequence is $\langle k : \mathcal{S}_j[k][O] = \mathcal{S}_j[i][O] \rangle$.
Since $O$ is unique, only $k = i$ satisfies this, so $w_i(\mathcal{S}_j) = \langle i \rangle$.

\item $P$ is a unique column combination and the frame does not exclude the current row.
Since $P$ uniquely identifies each tuple, each partition contains exactly one tuple.
Therefore, $|\mathcal{S}_j| = 1$ and $w_i(\mathcal{S}_j) = \langle 1 \rangle$.

\end{enumerate}
Empty frames occur when any of (i)--(iv) holds but the frame additionally excludes the current row, yielding $w_i(\mathcal{S}_j) = \langle \rangle$.

\subsection{Eliminating the Window Function}

Once the optimizer establishes that $|w_i(\mathcal{S}_j)| \leq 1$ for all $i,j$, \ie every frame contains at most one tuple, it must derive the scalar expression $e$ that replaces the window function $f$.
This derivation depends on whether $f$ operates on the frame or the partition.

Aggregate functions operate on the frame, but ranking and distribution functions, such as \texttt{row\_number()}, \texttt{rank()}, \texttt{dense\_rank()}, \texttt{percent\_rank()}, and \texttt{cume\_dist()}, operate on the entire partition.
In our example query \cref{lst:q-frame}, if a partition $\mathcal{P}_j$ had 10 rows, \texttt{row\_number()} still sees all 10 rows and returns values 1 through 10.
Thus, ranking and distribution functions reduce to constants only when the partition itself contains a single row, \ie when condition~(iv) holds.
\Cref{tab:frame-collapse} summarizes the replacement expressions and their applicability conditions.
\revisionA{R1.O3}{The cases with at most one tuple follow ordinary aggregation semantics, \eg \texttt{count} of an empty frame is 0.
The only exception is \texttt{count(a)} on a singleton frame and a nullable column, which needs \texttt{CASE WHEN a IS NULL THEN 0 ELSE 1 END}.}

\begin{table}[h]
\caption{Expression $\boldsymbol{e}$ replacing \revisionstyleA{\WF-specific} functions $\boldsymbol{f}$ for collapsing frames.}
\label{tab:frame-collapse}
\centering
\resizebox{\columnwidth}{!}{%
\begin{tabular}{@{}clcc@{}}
\toprule
$\boldsymbol{|\mathcal{W}|}$ & \textbf{Function $\boldsymbol{f}$} & \textbf{Expression $\boldsymbol{e}$} & \textbf{Conditions} \\
\midrule
$0$ & \texttt{count(a), count(*)} & 0 & any of (i)--(iv) \\
$0$ & any other & \textsf{NULL} & any of (i)--(iv) \\
\midrule
$1$ & \texttt{first\_value(a), last\_value(a)} & $a$ & any of (i)--(iv) \\
$1$ & \texttt{nth\_value(a, 1)} & $a$ & any of (i)--(iv) \\
$1$ & \texttt{nth\_value(a, n)}, $n \neq 1$ & \textsf{NULL} & any of (i)--(iv) \\
$1$ & \texttt{count(a)}, $a$ nullable & case expr. & any of (i)--(iv) \\
\midrule
$1$ & \texttt{rank(), dense\_rank(), row\_number()} & $1$ & (iv) \\
$1$ & \texttt{cume\_dist()} & $1$ & (iv) \\
$1$ & \texttt{percent\_rank()} & $0$ & (iv) \\
\bottomrule
\end{tabular}
}
\end{table}

Recall that condition (iv) implies single-row partitions.
Ranking functions assign rank 1 to the sole row.
The cumulative distribution function \texttt{cume\_dist()} returns $1/1 = 1$.
Function \texttt{percent\_rank()} is defined as $(\textit{rank} - 1) / (\textit{partition size} - 1)$; with one row, this is $0/0$, which the SQL standard defines as $0$.

\myparagraph{Implementation note}
Recognizing collapsing frames can already occur at parse time \revision{because} the conditions follow directly from translating the SQL 
frame clause to the internal representation.
Determining the replacement expression is a switch on $f$'s type or a lookup in a table like \cref{tab:frame-collapse}.

\section{Partition Analysis}
\label{sec:partition}
This section presents another reasoning technique that allows common optimizations to apply more broadly than previously assumed.
We start again with a motivating example.
Recall \cref{lst:q-partition} from the introduction.
The query computes a yearly average and filters to recent sales.
The predicate \texttt{sold\_date >= DATE'2020-02-15'} operates on the same column used in the partition expression \texttt{EXTRACT(YEAR FROM sold\_date)}.
One might hope to push the predicate through the window, but doing so directly would remove rows needed for the yearly average computation.
What can be done instead is pushing a \emph{derived predicate}, shown in gray in the following query:

\medskip
\begin{center}
\begin{tikzpicture}[
    every node/.style={anchor=west, inner sep=0pt, font=\ttfamily\footnotesize},
    node distance=0pt and 0pt
]
\node[font=\bfseries\footnotesize] (label) {\ref{lst:q-partition}$^\prime$:\setprimelabel{lst:q-partition}{lst:q-partition-prime}};
\node[right=0.2cm of label.east, anchor=west] (l1) {SELECT * FROM (};
\node[below=0.8\baselineskip of l1.west, anchor=west] (l2) {\hspace{2em}SELECT *, AVG(price) OVER (};
\node[below=0.8\baselineskip of l2.west, anchor=west] (l3) {\hspace{4em}PARTITION BY EXTRACT(YEAR FROM sold\_date)};
\node[below=0.8\baselineskip of l3.west, anchor=west] (l4) {\hspace{2em}) AS yearly\_avg};
\node[below=0.8\baselineskip of l4.west, anchor=west] (l5) {\hspace{2em}FROM sales};
\node[below=0.8\baselineskip of l5.west, anchor=west, minimum width=2em] (l6indent) {};
\node[right=0pt of l6indent.east, anchor=west, fill=gray!30, inner sep=2pt] (l6) {WHERE EXTRACT(YEAR FROM sold\_date) >= 2020) AS s};
\node[below=0.8\baselineskip of l6indent.west, anchor=west] (l7) {WHERE sold\_date >= DATE '2020-02-15'};
\end{tikzpicture}
\end{center}
\medskip

This transformation uses \cref{equiv:pd-pred2}.
Before describing when and how it derives such a predicate, we introduce the concept of partition integrity and work through simpler push-downs.

\subsection{Partition Integrity}
\label{ssec:integrity}

A window operator requires complete partitions to compute correctly.
Recall from \cref{ssec:formal} that partitions form in phase~1 and results collect in phase~6.
If an operator applied before phase~1 removes only some tuples from a partition, the window function computes over incomplete data, producing incorrect results.

The solution idea is simple: partitions must remain intact.
Functional dependencies can help determine when a predicate or join respects this constraint.
If an FD holds between the partition attributes $P$ and the filter or join attribute $b$, then all tuples in a partition share the same value for $b$.
A predicate on $b$ therefore either accepts every tuple in the partition or rejects all of them.
The same reasoning applies to join attributes.
\revisionB{R2.W1, R2.D1}{Such FDs are within reach of a modern optimizer: 
they follow from declared keys and uniqueness constraints, and can also be derived at optimization time, for instance, when an equality filter makes an attribute constant. 
Several optimizers already maintain functional dependencies for this kind of reasoning, as we discuss in \cref{sec:related}.}

The FD can run in two directions.
When $P \rightarrow b$, the partition determines the filter attribute, and we can push the predicate or join directly.
When $\{b\} \rightarrow p$ for any $p \in P$, the filter attribute determines a partition attribute.
We discuss each case in turn.

\subsection{Partition Determines Filter Attribute}
\label{ssec:p-implies-a}

\Cref{equiv:pd-pred1,equiv:pd-join1,equiv:pd-join2,equiv:pd-join3,equiv:pd-join4} require the FD $P \rightarrow b$.
When it holds, pushing the predicate directly preserves partition integrity.
When $P$ and $b$ are arbitrary expressions, checking whether this FD holds requires reasoning about expression semantics.
In the general case, this is undecidable~\cite{DBLP:books/daglib/0086373}.
However, we can identify a reduced subset of patterns for which the check is tractable:
\begin{enumerate}[label=(\roman*),leftmargin=*]
    \item $b \in P$: we expect most systems to use this variant already.
    \item \revision{Deterministic functions} $e(p), p \in P$: examples are arithmetic expressions with a constant, ($p+c; p\bmod c; \ldots$) or functions on $p$, such as substring or date component extraction.
    \item Hierarchical data types, such as dates and years, if the hierarchy is known: equi-joined attributes also apply because primary keys induce transitive FDs.
\end{enumerate}

An example of (ii) above would be if we inverted the filtering and partitioning expressions in \ref{lst:q-partition-prime},
\ie we partition by \texttt{sold\_date} but filter by year.
With the FD $\left\{\text{date}\right\} \rightarrow \left\{\text{year}\right\}$,
we could push down a predicate over year because it eliminates entire partitions.

\revisionB{R2.D2}{The join push-downs differ in cost. 
Semi- and anti-joins (\cref{equiv:pd-join3,equiv:pd-join4}) only remove tuples, so pushing them never increases the window's input. 
Inner and outer joins (\cref{equiv:pd-join1,equiv:pd-join2}) can expand it, so an engine should push them only when the join does not grow the input it feeds the window.}

\subsection{Filter Attribute Determines Partition}
\label{ssec:a-implies-p}

\Cref{equiv:pd-pred2} uses the FD in the other direction: $\{b\} \rightarrow p$ for some $p \in P$.
Unlike the previous case, we cannot push the original predicate directly.
Tuples in the same partition may have different values for $b$, so pushing would break partition integrity.

The key observation from \ref{lst:q-partition-prime} is that a bound on dates implies a bound on years.
Any date satisfying \texttt{sold\_date >= DATE '2020-02-15'} must fall in year 2020 or later.
We can therefore derive a weaker predicate, \texttt{EXTRACT(YEAR FROM sold\_date) >= 2020}, and push it instead.
We call the new constant in the derived predicate $c'$.
This eliminates entire partitions---years 2019, 2018, and earlier---before the window operator runs.
The original predicate remains above to filter tuples within the year-2020 partition.

Deriving $c'$ depends on the partition expression.
These three approaches cover common cases:
\begin{enumerate}[label=(\roman*)]
    \item For equality predicates and deterministic functions, such as \texttt{EXTRACT}, apply the partition function $e$ to $c$ directly, as in our example.
    \item For equality predicates and hierarchical data types (dates, times, etc.), use the hierarchy to derive the coarser bound.
    \item Use an order dependency $[b] \mapsto [e(b)]$ when the mapping preserves order for range predicates.
          Examples of such dependencies are, again, date and time component extractions or monotonic arithmetic expressions with a constant (addition, subtraction, multiplication, division, but not modulo).
          Exclusive range predicates ($<, >$) must be transformed to their inclusive equivalent to preserve partition integrity.
\end{enumerate}

\myparagraph{Implementation Note}
Using (i) above, an engine would need to evaluate an expression $c^\prime = e(c)$ dynamically during optimization time. 
We expect, however, that $e$ would already be parsed, and therefore, the evaluation to be relatively fast because no table data needs to be accessed.
Additional machinery would be required to inspect expression
$e$ and recognize, from its shape, that it contains a function covered by case (i) above.

\section{Co-Evaluation}
\label{sec:coevaluation}
Partition Analysis covered predicates that can be pushed through the window operator.
Some predicates, however, depend on the window's output and cannot be moved as above.
Rather than evaluating these predicates after the window completes, we can push aspects of them into the window's internal execution, reducing data early.
As we mentioned above, we call this technique Co-Evaluation.

\subsection{Motivation}
\label{ssec:co-eval_motivation}

\Cref{equiv:result-limit} shows that a predicate on a ranking function can become a \texttt{LIMIT} when the window has no partitions and the sort list is a unique column combination.
Here is an example of a query that can benefit, and its restructured version:

\medskip
\noindent\hspace{0.4em}%
\begin{tikzpicture}[
    every node/.style={anchor=west, inner sep=0pt, font=\ttfamily\footnotesize},
    node distance=0pt and 0pt
]
\node[font=\bfseries\footnotesize] (label1) {\refstepcounter{exquery}\theexquery:\label{lst:q-ucc}};
\node[right=0.5cm of label1.east, anchor=west] (l1) {SELECT * FROM (};
\node[below=0.8\baselineskip of l1.west, anchor=west] (l2) {\hspace{2em}SELECT *, RANK() OVER (};
\node[below=0.8\baselineskip of l2.west, anchor=west] (l3) {\hspace{4em}ORDER BY <UCC expr> DESC) AS rnk};
\node[below=0.8\baselineskip of l3.west, anchor=west] (l4) {\hspace{2em}FROM r) AS s};
\node[below=0.8\baselineskip of l4.west, anchor=west] (l5) {~WHERE rnk <= 3};
\end{tikzpicture}
\medskip

\medskip
\begin{center}
\begin{tikzpicture}[
    every node/.style={anchor=west, inner sep=0pt, font=\ttfamily\footnotesize},
    node distance=0pt and 0pt
]
\node[font=\bfseries\footnotesize] (label2) {\ref{lst:q-ucc}$^\prime$:\label{lst:q-ucc-prime}};
\node[right=0.5cm of label2.east, anchor=west] (l6) {SELECT * FROM (};
\node[below=0.8\baselineskip of l6.west, anchor=west] (l7) {\hspace{2em}SELECT *, RANK() OVER (};
\node[below=0.8\baselineskip of l7.west, anchor=west, minimum width=4em] (l8indent) {};
\node[right=0pt of l8indent.east, anchor=west, fill=gray!30, inner sep=2pt] (box1) {ORDER BY <UCC expr> DESC)};
\node[below=0.8\baselineskip of l8indent.west, anchor=west] (l9) {\hspace{2em}FROM r)};
\node[below=0.8\baselineskip of l9.west, anchor=west, fill=gray!30, inner sep=2pt] (box2) {\sout{ORDER BY <UCC expr> DESC}};
\node[below=0.8\baselineskip of box2.west, anchor=west, fill=gray!30, inner sep=2pt] (box3) {LIMIT 3};
\node[right=0.5cm of box1.east, anchor=west, yshift=-2.0\baselineskip, font=\sffamily\footnotesize] (pushlabel) {push into the window};
\draw[thick, -{Stealth}] (pushlabel.west) -- (box1.east);
\draw[thick, -{Stealth}] (pushlabel.west) -- (box3.east);
\end{tikzpicture}
\end{center}
\medskip

The query ranks rows by a unique expression and keeps only the top three.
This pattern, ranking followed by a threshold filter, is common in analytics, for example when finding best-performing items or most recent entries.
The restructured version replaces the rank filter with a \texttt{LIMIT}\@.
This works because no ties can occur without partitioning and a unique ordering.
The first three rows in sort order are exactly those with rank at most three.
The proof sketch makes this precise.

\myparagraph{Proof sketch for \cref{equiv:result-limit}}
With $P = \varnothing$, all tuples belong to a single partition ordered by $O$ (phase~2).
Function \texttt{row\_number()} assigns increasing integers along this order.
Functions \texttt{rank()} and \texttt{dense\_rank()} do so as well when $O$ is a unique column combination.
Filtering to ranks at most $c$ therefore selects exactly the first $c$ tuples in sort order, which is equivalent to $\lambda_c(\tau_O(\cdot))$. \hfill$\square$
\smallskip

Since the sort is duplicated in the restructured version, an optimizer could conceivably push the LIMIT from the outer sort into the window's internal sort, pruning data early in the query and eliminating the outer sort entirely.

This is a powerful optimization.
However, it requires the \WF to not have partitioning.
Our motivation query \Cref{lst:q-coeval} partitions over \texttt{dept}, so \cref{equiv:result-limit} does not apply.
This is where the notion of Co-Evaluation may help.

\subsection{MOD Predicates}
\label{ssec:pushing_limit}

Any valid predicate over a window function can be co-evaluated while the function result is computed.
Rather than adding another pass on the data to calculate the predicate, an engine can do so in phase 4 of the \WF definition (\cref{ssec:formal}).
This will save the work of going over the data once more after the \WF calculation \revision{finishes}.

Some predicates, however, present a \emph{stopping property}.
Recall that phase~2 of the \WF definition (\cref{ssec:formal}) sorts each partition into a sequence $\mathcal{S}_j$.
With respect to this order, a predicate $C$ over a window function may, once it becomes false, stay false throughout the rest of the sequence.
More formally:
\[
\forall i \in \{1, \ldots, |\mathcal{S}_j|-1\}: \neg C(t_i) \Rightarrow \neg C(t_{i+1})
\]
We call a predicate that \revision{presents} this property \emph{monotonic order-dependent} (MOD).
In \cref{lst:q-coeval}, the \texttt{rnk <= 3} predicate is MOD.
A running sum over non-negative values with a cumulative frame (\eg \texttt{ROWS BETWEEN UNBOUNDED PRECEDING AND CURRENT ROW}) is also MOD.
Once the predicate \texttt{SUM(a) OVER (...) < threshold} becomes false, it remains false for the rest of $\mathcal{S}_j$.
The same intuition holds for the time interval elapsed since a fixed point.
\revisionB{R2.W3, R2.D3}{While a between predicate on the \WF result is not strictly MOD, the predicate's upper bound still implies a stop criterion that can be used for pruning.}
\revisionD{Meta}{In the case study we mentioned in \cref{sec:intro}, 8~of the 10~most expensive \WF queries featured a MOD predicate.}

\subsection{Pruning in Early Phases of a \WF}

\Cref{equiv:co-eval} enables two optimizations based on MOD predicates.
The first is \textit{early stop}, where we can iterate through the partition in sort order only until the predicate turns false.
For predicates that tend to turn false after just a few elements, such as rank with low $k$, this can save many iterations.
This raises a question: why build and sort a full partition if we will not traverse it completely?

The alternative is to prune rows that cannot qualify even before sorting.
We call this \textit{early pruning}.
In our example, rows that cannot rank in the top $k$ can be discarded during the partitioning process, as we show shortly.

Early pruning can deliver substantial performance improvements by reducing the data that reaches the sort.
From \revision{a} complexity standpoint, if $n$ tuples are distributed across $m$ partitions, sorting costs $\landO(\frac{n}{m} \log \frac{n}{m})$ per partition. 
Conceptually, maintaining a priority queue of size $k$ for each partition suffices to answer the query without sorting when $k$ is small.
This costs $\landO(\frac{n}{m} \log k)$ per partition, or $\landO(n \log k)$ overall.
When $k \ll \frac{n}{m}$ (as is typical for top-$k$ queries), the savings are substantial.
For \texttt{rank()} and \texttt{dense\_rank()}, the queues must handle duplicates and have a size $k'\geq k$, unless a UCC on the ordering attributes ensures no ties occur.

Applying these ideas to \cref{lst:q-coeval} changes its \WF execution as shown below, with the clauses involved in pruning highlighted:

\medskip
\begin{center}
\begin{tikzpicture}[
    every node/.style={anchor=west, inner sep=0pt, font=\ttfamily\footnotesize},
    node distance=0pt and 0pt
]
\node[font=\bfseries\footnotesize] (label) {\ref{lst:q-coeval}$^\prime$:\setprimelabel{lst:q-coeval}{lst:q-coeval-prime}};
\node[right=0.5cm of label.east, anchor=west] (l1) {SELECT * FROM (};
\node[below=0.8\baselineskip of l1.west, anchor=west] (l2) {\hspace{2em}SELECT *, RANK() OVER (};
\node[below=0.8\baselineskip of l2.west, anchor=west, minimum width=4em] (l3indent) {};
\node[right=0pt of l3indent.east, anchor=west, fill=gray!30, inner sep=2pt] (partition) {PARTITION BY dept};
\node[below=0.8\baselineskip of l3indent.west, anchor=west, minimum width=4em] (l4indent) {};
\node[right=0pt of l4indent.east, anchor=west, fill=gray!30, inner sep=2pt] (orderby) {ORDER BY salary DESC};
\node[right=0pt of orderby.east, anchor=west] (l4rest) {) AS rnk};
\node[below=0.8\baselineskip of l4indent.west, anchor=west] (l5) {\hspace{2em}FROM employees) AS s};
\node[below=0.8\baselineskip of l5.west, anchor=west, fill=gray!30, inner sep=2pt] (l6) {WHERE rnk <= 3};
\node[right=1.5cm of l6.east, anchor=west, font=\sffamily\footnotesize] (modlabel) {MOD predicate, $k = 3$};
\draw[thick, -{Stealth}] (modlabel.west) -- (l6.east);
\node[right=0.7cm of partition.east, anchor=west, yshift=2\baselineskip, font=\sffamily\footnotesize] (prunelabel) {prune to top-$k$ per partition};
\draw[thick, -{Stealth}] (prunelabel.west) -- (partition.east);
\draw[thick, -{Stealth}] (prunelabel.west) -- (orderby.east);
\end{tikzpicture}
\end{center}

\myparagraph{Implementation note}
Recognizing MOD predicates can occur at optimization time by inspecting the predicate structure and the window function type.
Applying the optimizations, however, requires operator modifications.
We call such operators \WFPLUS and describe one implementation next.

\section{\WFPLUS Implementation}
\label{sec:implementation}
\begin{figure}[ht]
\centering
\resizebox{\linewidth}{!}{\input{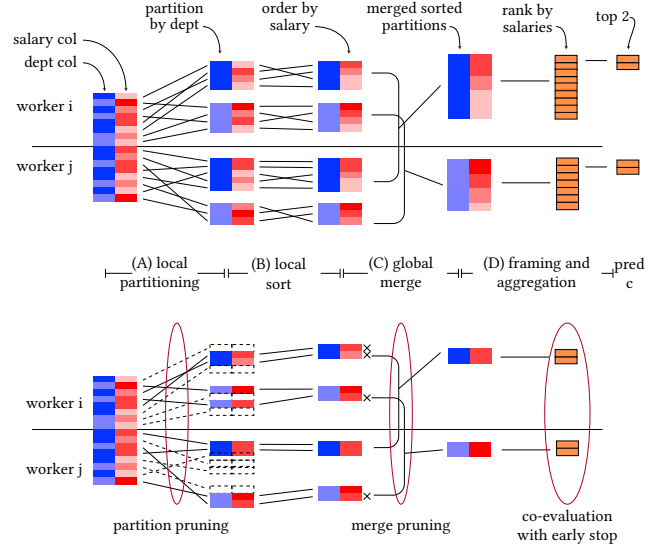}}
\caption{DuckDB's \WF pipeline for \cref{lst:q-coeval}.
We use \texttt{rnk<=2} for the purpose of this illustration.
Top: standard execution. Bottom: \WFPLUS with pruning at different phases.
The benefits of \WFPLUS's pruning points \revision{are} cumulative, as early pruning reduces the work of all subsequent phases.}
\label{fig:wfplus_execution}
\end{figure}

The phases in \cref{sec:semantics} reflect how window functions may be implemented in commercial systems.
Changing some of these phases to perform early pruning can, therefore, transform an existing operator into \revision{a} \WFPLUS one.
To show how, we integrated our techniques into DuckDB, a modern engine geared towards analytical queries~\cite{DBLP:conf/sigmod/RaasveldtM19} and contrast its native execution and our changes in \Cref{fig:wfplus_execution}.

As the top of the figure shows, DuckDB's native \WF is a parallel implementation that follows \citet{DBLP:journals/pvldb/LeisKK015}: 
(A)~local partitioning distributes tuples from morsels (\emph{chunks}) across local radix buckets  \revisionA{R1.O3}{by hashing each tuple's partition key, so a single window partition may span several buckets and a bucket may hold several partitions}, 
(B)~local sort orders tuples within each worker's buckets, 
(C)~global multi-way merge combines sorted runs~\cite{DuckDBSort,DBLP:conf/icde/KuiperM23} \revisionstyleA{so that workers scale without sorting the entire relation at once}, and 
(D)~framing/aggregation computes the results\revisionstyleA{, using a segment tree to share aggregation work across overlapping frames}.
The bottom part of the figure depicts three pruning points where \WFPLUS techniques can reduce data volume.
\revisionA{R1.O3}{Note that \WFPLUS does not exit or skip any of these phases. As \Cref{fig:wfplus_execution} shows, every phase still runs, and the pruning points only reduce the data volume that reaches subsequent phases.}

\subsection{Pruning Points}

We next describe these optimizations, going from late to early steps.

\myparagraph{Co-Evaluation with Early Stop} 
We implement Co-Evaluation on the framing and aggregation step. 
It decides if a tuple matches a predicate while iterating the materialized chunks generated by the merge phase. 
This implementation can be used for any predicate, but it is most beneficial when the predicate involves the \WF result.
If the co-evaluated predicate is MOD, we skip the remaining partition elements once the predicate turns false.
As mentioned above, using MOD predicates to prune on or before the sort yields better results.

\myparagraph{Merge Pruning} 
It is possible to take MOD predicates into consideration during the merge step depicted in \Cref{fig:wfplus_execution}. 
In our implementation, we extend the sorted-batch merger to do so. 
It stops reading tuples from (initial and merged) sorted runs beyond their rank-$k$ element for each partition. 
We only consider ranking, but this strategy can conceptually generalize to further MOD predicates at higher evaluation cost.
Our merger takes~$k$ as input, but that is the extent of state added at this pruning point. 
Merge pruning does not add any overhead other than additional key comparisons. 
As expected, materialization, framing, and the \WF evaluation benefit from less data read from the result of this step.

\myparagraph{Partition Pruning} This is the earliest pruning point.
Note that a bucket at this point does not observe all of the query's partition key values. 
A partition may be scattered across buckets of several workers (as \Cref{fig:wfplus_execution} shows), or a bucket may contain several partitions (not shown).
In our implementation, we dynamically create heaps of size $k$ for each partition per worker.
When a tuple gets assigned to a bucket, we identify its partition.
If the ordering value is larger than the respective heap's maximum value, we disregard the tuple.
Otherwise, we update the heap.
For \cref{lst:q-coeval}, elements outside the top-3 salaries per department can get discarded here.

The benefits of pruning data this early can be significant.
With $m$ \WF partitions, the sort phase reads only $m \cdot k$ instead of all $n$ tuples in the best case, reducing sort complexity from $\landO(n  \cdot \log n)$ to $\landO(m \cdot k \cdot \log (m \cdot k))$.
This improvement is immense if there are only a few but large partitions and $k$ is small.
However, in the worst case, all partitions have too few tuples, and each worker reads fewer than $k$ tuples per \WF partition, but still inserts them into the heaps.
The overhead of iterating over the tuples to filter them and maintaining $k$-sized heaps can grow to dominate the cost  (see \cref{sec:experiments:ablation}).

\subsection{Adaptive Partition Pruning}
\label{ssec:adaptive_pruning}

To mitigate this case during execution, we introduce an adaptive partition pruning strategy.
Its main idea is driven by the following question: 
\emph{how many tuples are required to answer the query?}
A query that filters for the top-$k$ tuples of $m$ partitions and processes more than $m \cdot k$ tuples can potentially discard excess ones.
Put differently, if $m \cdot k \cdot \delta \mathbin{\revision{<}} \vert \text{read tuples} \vert$, pruning could theoretically disregard tuples for $\delta \mathbin{\revision{\geq}} 1$.
In practice, the overhead of the heap maintenance may still dominate the cost if $\delta$ is only marginally over~1.
We have determined experimentally that $\delta$ should be 5 (see \Cref{sec:experiments:threshold}).
This value seems high at first, but it is still a linear factor compared to linearithmic sorting, which Adaptive Partition Pruning can shorten.

The threshold, however, is only part of the problem.
The workers we see in \Cref{fig:wfplus_execution} do not know $m$, the number of partitions, in advance.
In fact, the precise $m$ will not be known until the end of the query's execution.
Thus, we estimate $m$ \emph{on a per\revision{-}worker basis.}
For each chunk, a worker scans \revisionA{R1.O3}{all} tuples and calculates a hash for the partition value \revisionA{R1.O3}{before partitioning the data}.
Our \WFPLUS implementation uses that step to also update an instance of a HyperLogLog sketch~\cite{DBLP:conf/edbt/HeuleNH13}.
The sketch can provide a good approximation $m'$ with little overhead and no additional statistics or previous optimizer decision.
With this up-to-date estimation, we decide whether to prune for each chunk individually \revisionA{R1.O3}{and upfront} by calculating $m' \cdot k \cdot \delta$ and comparing that value to the number of tuples the worker read so far.

\revisionB{R2.D2}{This makes co-evaluation's cost a runtime decision rather than an optimizer one. 
Each worker decides per chunk from its HyperLogLog estimate and the threshold $\delta$, so the operator captures the cost-dependent case at execution time without a static cost model.}

\section{Evaluation}
\label{sec:experiments}
The experimental evaluation of the presented techniques is divided as follows:
\cref{sec:experiments:analyses} evaluates the equivalences;
\cref{sec:experiments:ablation} presents an ablation study of our \WFPLUS operator's pruning points;
\cref{sec:experiments:skew} expands that study to skew; 
\cref{sec:experiments:threshold} experimentally determines $\delta$; 
and \cref{sec:evaluation:pipeline} analyses early pruning in detail.

\myparagraph{Setup}
We execute the microbenchmarks on one Intel Xeon Platinum~8180 socket 
with \qty{378}{GiB} of NUMA-local memory.
For stable results, we limited execution to the 28~physical cores.
The machine is running Ubuntu~24.04, and we compile DuckDB and our \WFPLUS implementation over it with \revision{GCC-14.2} and \texttt{-O3}.

\myparagraph{Methodology and Datasets}
All our experiments execute actual queries in DuckDB unless stated otherwise.
In some experiments, we have baseline and optimized queries to compare.
In other experiments, we compare the official distribution of DuckDB with our modified one.
The data we used in the experiments consists of synthetic relations varying in size, number of partitions, and skew.
To eliminate any variation in response time, we execute each experiment from 10 to 1000 times and report the median runtime.

\subsection{Equivalence Microbenchmarks}
\label{sec:experiments:analyses}

\begin{figure*}[tb]
\centering
\includegraphics[width=\textwidth]{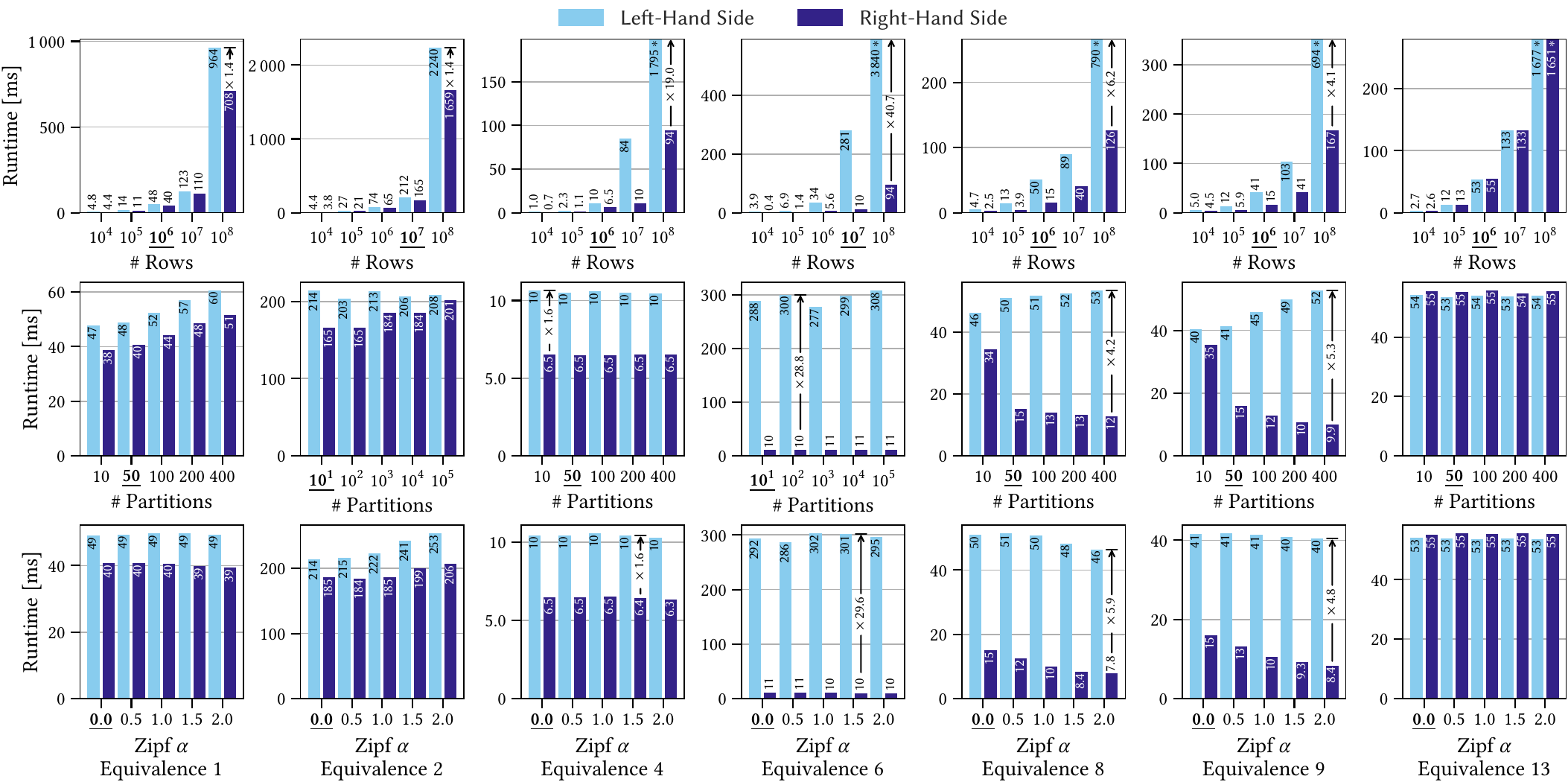}
\caption{%
\revisionstyle{Runtime of each equivalence's left-hand side (native execution) vs.\ right-hand side (with our optimization), across varying table size, partition count, and skew. 
Each column corresponds to one equivalence class from \Cref{tab:transformations}.
Within each column, parameter values that are fixed in the other charts are \underline{marked}.
Speedup factors are annotated where applicable.}%
}
\label{fig:primary_expr}
\end{figure*}

\revisionM{R1.O2, R3.W1, R3.W2, R3.D1}{To evaluate the benefits of the analyses discussed in \cref{sec:frame,sec:partition}, and to cover the full equivalence table, we execute a before-and-after query pair for each equivalence class in \Cref{tab:transformations}.
The left-hand side of each pair presents the pattern on the left-hand side of an algebraic equivalence of the table, and the right-hand side applies that equivalence’s rewrite, just as in~\cref{lst:q-frame} vs.\ \ref{lst:q-frame-prime}.
We study three facets per equivalence, varying the number of rows in the base table, varying the number of partitions at a fixed table size, and varying the skew ($\alpha$ in a Zipf distribution) at fixed size and partition count.
The results appear in \Cref{fig:primary_expr}.
Each equivalence employs a table specifically designed for its scenario, resulting in different sweep ranges for each column in the figure.
}

\revisionstyle{We observe that our variations match or outperform the original queries in every tested case. 
The frame-collapse equivalence (\cref{equiv:replace-projection}) yields the largest gains, up to 40.7$\times$ on the largest table, because it removes the window operator entirely. 
The push-down equivalences improve as the number of partitions or the degree of skew grows because aligning the predicate with the partition expression discards entire partitions before the window runs. 
The remaining equivalences deliver consistent speedups by shrinking the sort and partition keys. 
In the few cases where an equivalence offers no benefit, its rewrite stays on par with the original query rather than degrading it.}

\subsection{Co-Evaluation and its Optimizations}
\begin{figure*}[tb]
    \centering
    \includegraphics[width=\textwidth]{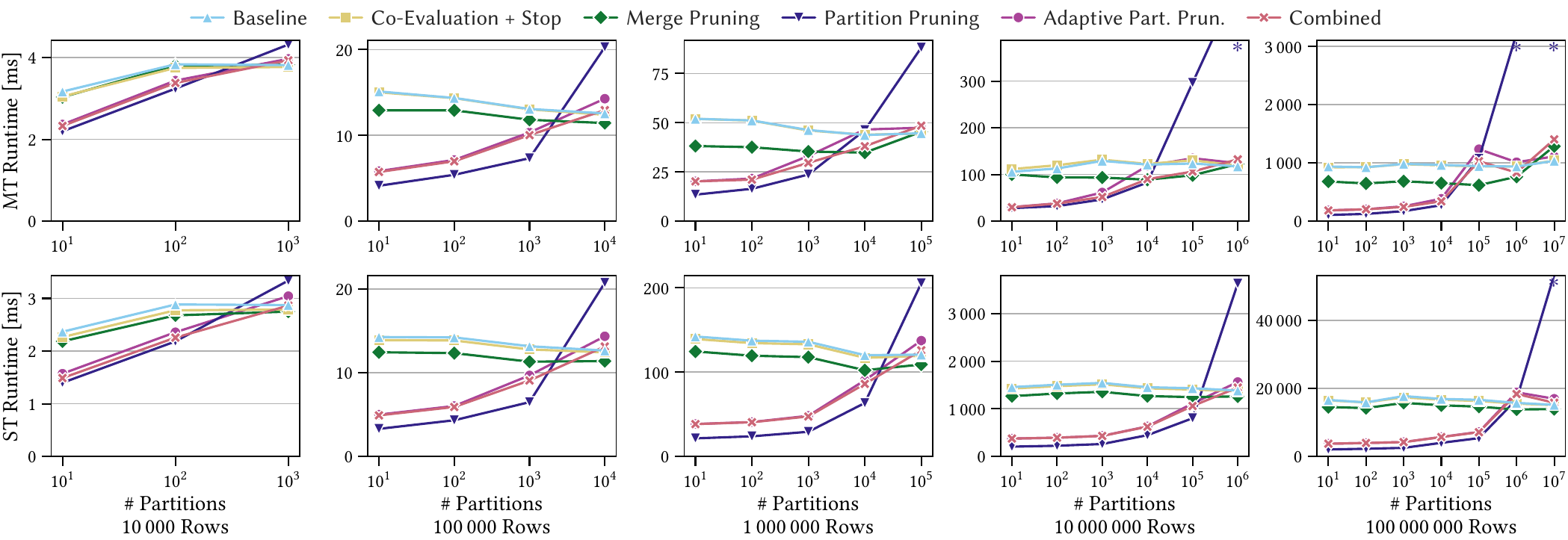}
    \caption{Microbenchmarks for Co-Evaluation (\cref{lst:q-coeval}, $\boldsymbol{k=3}$, $\boldsymbol{\delta = 5}$) using different optimizations for different relation sizes and numbers of \WF partitions, multi- (MT) and single-threaded (ST). Asterisk $\ast$ depicts a time-out within three times of the Baseline.
    The combination of optimization techniques is almost always considerably faster than the baseline.}
    \label{fig:evaluation:microbenchmarks-unskewed}
\end{figure*}

To evaluate the benefits of the pruning techniques from \Cref{sec:implementation}, we measure the performance of our \WFPLUS implementation in isolation.
We test each pruning technique individually and contrast it with DuckDB's unaltered \WF implementation. 
We also experiment with combining all the techniques at once.
\label{sec:experiments:ablation}
\Cref{fig:evaluation:microbenchmarks-unskewed} shows multi-threaded (MT) and single-threaded (ST) runtimes for a query resembling \cref{lst:q-coeval} on uniformly distributed partitions.

Co-Evaluation with Early Stop does not show significant improvements over the baseline.
Though this technique reduces the tuples emitted by the operator, all tuples have been partitioned, sorted, merged, and materialized into chunks before.
Thus, the optimization potential is small for DuckDB's execution model, highlighting the necessity for optimizations at earlier pipeline stages.

Merge Pruning shows consistent improvements as long as it can effectively reduce the \revision{number} of tuples:
it is around 1.2$\times$ to 1.6$\times$ faster than the baseline for large relations and few partitions.
For the numerous small partitions case, the local sorted runs do not contain enough tuples per \WF partition for pruning, and merge pruning can only reduce the tuples that are materialized, framed, and aggregated.
This is the case for the right-most data point in the MT plots above \num{100000}~rows: DuckDB uses multiple workers that only see few (if any) of the ten tuples per partition.
Here, the performance improvement diminishes.

Partition Pruning is extremely beneficial for large partitions, where many tuples can be aggressively discarded early.
We observe a maximum 9$\times$ speedup for \qty{100}{million} rows and 10~partitions.
However, the bookkeeping overhead grows and dominates the benefit, at \revision{least} when few or no tuples can be discarded from local partitions.
In the worst cases, the performance completely degrades and hits the timeout of 3$\times$ the baseline runtime.
For the datasets with 10 and \qty{100}{million} tuples, DuckDB utilizes all \num{28}~workers, so this tipping point is already reached with a partition size of \num{100}~tuples.
This behavior shows that partition pruning requires our adaptive approach to be feasible in practice.

Adaptive Partition Pruning detects unfavorable cases, switching to regular execution so performance never drastically drops when partition pruning produces considerable overhead.
There is only one case (\qty{10}{million} rows, \num{10000}~partitions) where it switches too early; in the remaining cases, it closely matches regular partition pruning with 2$\times$--5$\times$ speedups.
The difference compared to the regular version arises from HyperLogLog maintenance for every chunk, which could be reduced by fewer updates.
Overall, this heuristic makes partition pruning viable for different dataset characteristics.

Lastly, we combine Co-Evaluation, Merge Pruning, and Adaptive Partition Pruning for the final \WFPLUS implementation.
The runtimes are close to or even below the best of the individual techniques and show that this combination is ultimately beneficial:
in most of the cases where Merge Pruning or Adaptive Partition Pruning outperform each other, we benefit from the best of both.
The performance is drastically better than the baseline for few and large partitions, and only converges to the baseline for really small partitions. 

\subsection{Skewed Data}
\label{sec:experiments:skew}
\begin{figure}[tb]
    \centering
    \includegraphics[width=\columnwidth]{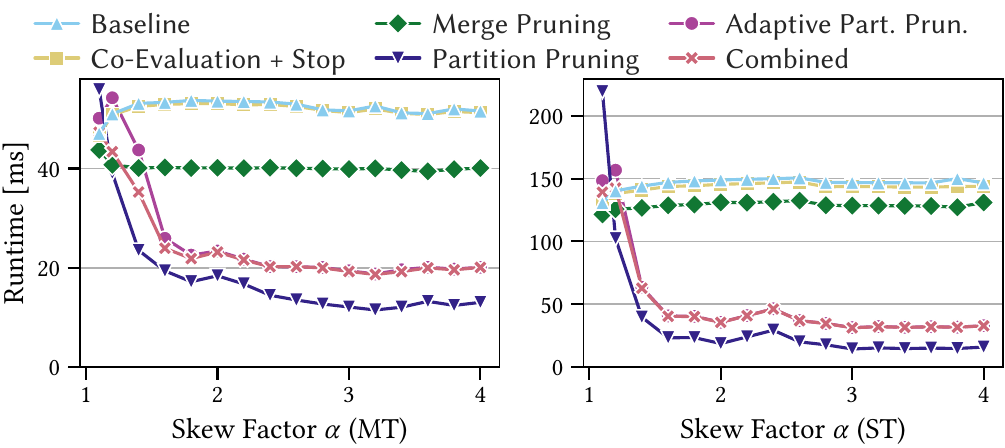}
    \caption{Microbenchmarks for Co-Evaluation (\cref{lst:q-coeval}, $\boldsymbol{k=3}$, $\boldsymbol{\delta = 5}$) using different optimizations for \num{1000000} tuples and different skew factors of a zeta distribution, multi- (MT) and single-threaded (ST).}
    \label{fig:evaluation:microbenchmarks-skewed}
\end{figure}

Real-world data is never uniformly distributed~\cite{DBLP:journals/tkde/ZhangR22}, so we modify the setup of \cref{sec:experiments:ablation} by varying the degree of skew.
We fix the relation size to \qty{1}{million} tuples and distribute the partitions according to a zeta distribution with a skewness factor $\alpha$ between 1.1 and 5.
Higher $\alpha$ corresponds to 
a larger share of data that concentrates on a few partitions.
\Cref{fig:evaluation:microbenchmarks-skewed} shows the results of this experiment.

The proposed optimizations perform well also on skewed datasets.
As in our previous experiments, Merge Pruning consistently achieves 1.2$\times$ to 1.5$\times$ speedups compared to the baseline in most cases.
Partition Pruning speeds up the execution up to a factor of 5~(MT) or 10~(ST) for high skew factors/few partitions because most of the large partitions' tuples can be discarded.
However, there is some overhead for many small partitions.
Adaptive Partition Pruning does not directly pick to prune for many partitions, but soon reaches performance close to the regular version.
Again, the combination of optimizations almost always achieves the best performance of the individual approaches, with speedups around 3$\times$ (MT) and 5$\times$ (ST).

\subsection{Adaptive Partition Pruning Threshold}
\label{sec:experiments:threshold}

\begin{figure}[tb]
    \centering
    \includegraphics[width=0.45\columnwidth]{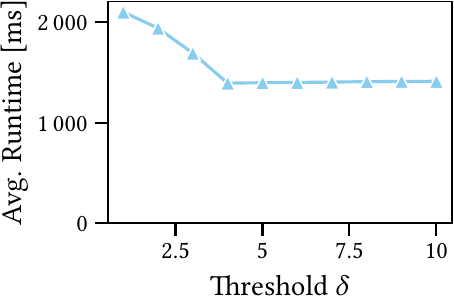}
    \caption{Average runtimes for adaptive partition pruning with different $\boldsymbol{\delta}$ across all configurations seen in \cref{fig:evaluation:microbenchmarks-unskewed}.
    }
    \label{fig:evaluation:thresholds-unskewed}
\end{figure}

\Cref{sec:implementation} described the Adaptive Partition Pruning strategy and introduced the threshold $\delta$ as a tuning parameter.
To assess its influence and find a value that performs well for different scenarios, we execute all configurations from \cref{sec:experiments:ablation} with Adaptive Partition Pruning and $\delta$ between 1 and 10.
Then, we calculate the average runtime across all configurations, shown in \cref{fig:evaluation:thresholds-unskewed}.

While the concrete value for $\delta$ may vary for individual \WFPLUS implementations, this experiment demonstrates that such a threshold exists.
In general, higher values perform better, and the aggregated runtimes vary by a factor of 1.5 between the best and the worst value for $\delta$.
Small thresholds cannot detect cases where the overhead of partition pruning deteriorates.
Too large a value stops pruning too early when it is still beneficial.
In our experiments, $\delta = 5$ is the optimal value, and the more detailed evaluation in \cref{sec:experiments:ablation} shows that it performs well in most of the cases.

\subsection{Detailed Pipeline Analysis}
\label{sec:evaluation:pipeline}

\begin{figure}[tb]
\centering
\includegraphics[width=\columnwidth]{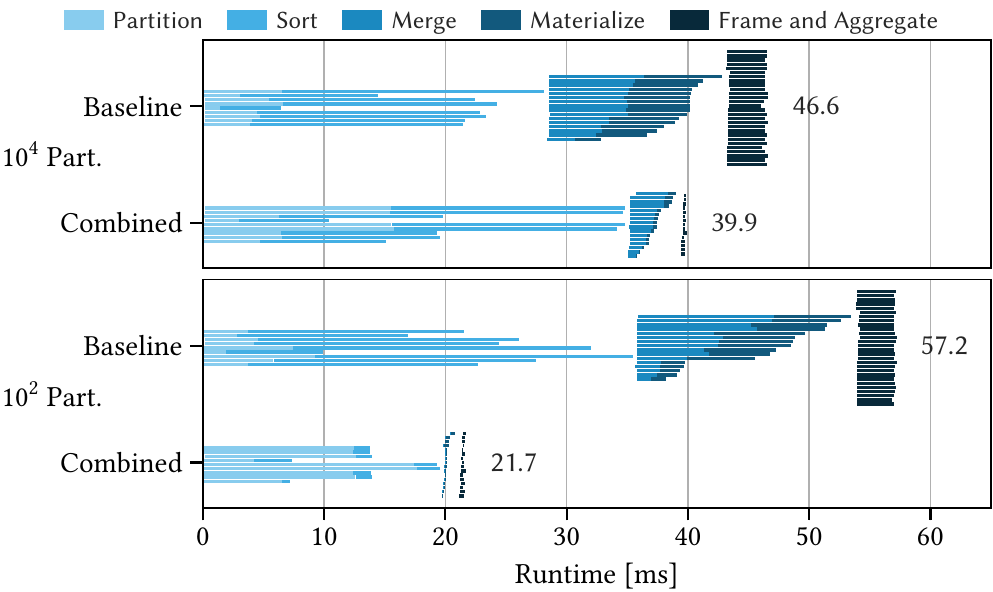}
\caption{Per-worker execution timeline for different \WF processing phases (cf.~\Cref{fig:wfplus_execution}) for~\cref{lst:q-coeval} and $\boldsymbol{k=3}$ on \qty{1}{million} rows.  
Each line represents a worker and each phase may use a different number of workers.
Combined pruning reduces total time by eliminating later work in early phases.}
\label{fig:evaluation:workers}
\end{figure}

In this last experiment, we investigate the impact of Co-Evaluation with combined optimizations on individual pipeline steps of the \WF evaluation for \cref{lst:q-coeval}.
We use a dataset of \qty{1}{million} rows and two partition counts $m_1=100$ and $m_2=\num{10000}$.
For all configurations, DuckDB uses 9~workers for partitioning and local sort.
These workers radix-partition the tuples into 16~local radix buckets, and all related local radix buckets are merged to a single sorted bucket and materialized by one worker per global bucket.
Finally, 25~workers perform framing and the window function computation.
\Cref{fig:evaluation:workers} shows the timeline for each phase per worker.

At a high level, we observe that Adaptive Partition Pruning improves performance by a factor of almost 3 for 100~partitions, and that Merge Pruning is beneficial for \num{10000}~partitions. 
For 100~partitions, local partitioning takes longer than the baseline due to Partition Pruning. Workers maintain a HyperLogLog and partition-local heaps to reject tuples outside the top-3.
As a result, radix buckets are much smaller, and local sorting is much faster.
Only $m_1 \cdot k = 300$~tuples must be read per worker and from the merged buckets.
Thus, merging, materialization, and framing/aggregation happen almost instantly compared to the baseline.
The additional overhead in partitioning pays off in all subsequent phases.

The performance benefit of optimized Co-Evaluation is not as large for \num{10000}~partitions.
Here, Adaptive Partition Pruning introduces overhead with the HyperLogLog, as the workers decide against partition pruning.
Thus, the local sort times are comparable to the baseline.
With Merge Pruning, we need to read $m_2 \cdot k = \num{30000}$~tuples from each local run, and merge \num{270000}~tuples overall.
This number is higher than for 100~partitions, but still considerably lower than in the unmodified operator, which reflects in the merge runtime.
Merge Pruning continues to reduce the number of tuples for materialization and framing/aggregation:
it is a good fallback strategy when Adaptive Partition Pruning cannot eliminate tuples.

\section{Related Work}
\label{sec:related}
To the best of our knowledge, ours is the first work to provide a rigorous table of equivalences involving \WFs.
The work, however, intersects with several efforts in database query optimization.
We present these intersections in groups.

\myparagraph{\revisionstyleB{Dependency-aware optimization}}
\revisionB{R2.D1}{The reductions and push-downs in \cref{tab:transformations} rely on functional and order dependencies, which modern optimizers increasingly maintain. 
Apache Calcite~\cite{DBLP:conf/sigmod/BegoliCHML18} tracks functional dependencies for query rewriting, and Orca~\cite{DBLP:conf/sigmod/SolimanAREGSCGRPWNKB14}, the optimizer behind Greenplum, represents them natively. 
\citet{DBLP:conf/sigmod/SimmenSM96} showed how order properties are derived and propagated during optimization.
\citet{DBLP:journals/vldb/KossmannPN22} surveyed about 60 optimization techniques that use data dependencies.
Obtaining and exploiting further dependencies for window optimization is an active direction, including work by some of the authors~\cite{DBLP:conf/edbt/LindnerRN26}.}

\myparagraph{Previous \WF treatments}
The approach most similar to ours appears in \citet{BellamkondaBozkayaGuptaEtAl2000}, who presented techniques such as merge pruning.
That paper, however, missed other earlier-stage pruning opportunities that we present here.
Moreover, the absence of a clear definition of \WFs makes it difficult to assess the correctness of the proposed query manipulations.
Regarding the works we cited in~\Cref{sec:intro}, as we mentioned there, our intention is to complement them and provide a rigorous base for others to do so as well.
In particular, \citet{DBLP:journals/pvldb/CaoCLT12} proposed that cover-set search can be improved with the kind of functional dependency usage of \cref{equiv:reduce-order1,equiv:reduce-order2}.
\revisionC{R3.W3, R3.D2}{Cover-set search is precisely the joint optimization of multiple windows that share a partitioning or ordering while computing different aggregates over the same sorted data. Our equivalences are complementary, as the reductions can shrink the keys such windows share before the sharing takes place.}

\myparagraph{Nested relational algebra}
Since the foundational work of \citet{DBLP:conf/vldb/Makinouchi77}, \citet{DBLP:conf/pods/JaeschkeS82}, and \citet{DBLP:journals/tods/RothKS88}, different authors have used nested relations and nested algebra to reason about SQL structures.
\citet{DBLP:conf/cikm/LiuR94} and \citet{DBLP:journals/is/LiuY05} defined algebraic equivalences.
\citet{DBLP:journals/tods/CaoB07} used nested algebra to optimize subqueries.
\citet{DBLP:conf/sigmod/AuerbachHMSS17} provided an extension, \eg to model views.
However, at that time, those works did not consider ordered tuple sequences.

\section{Conclusion}
\label{sec:conclusion}
\revision{In this work, we introduced a table of algebraic equivalences for window functions that formalizes existing transformations and introduces novel ones. 
We showed that certain transformations are applicable to a broader range of cases when specific conditions regarding window framing and partitioning are considered. 
Notably, one transformation enables the early application of a window predicate, which is typically executed at the conclusion of the window function, by pushing it to earlier stages. 
This optimization, referred to as \WFPLUS, allows for data pruning during the initial phases of window function execution. 
Collectively, these transformations, along with the \WFPLUS optimization, yielded significant performance improvements across a wide variety of tested scenarios.}

\begin{acks}
We thank Martin Boissier and Matthias Herzog for their assistance with the case study and Sarah Kleest-Meißner for her feedback on the manuscript.

Daniel Lindner received funding from SAP~SE.
\noindent{\raisebox{-0.15\height}{\euflag}} This project was partially funded by the European Union. Views and opinions expressed are however those of the author(s) only and do not necessarily reflect those of the European Union or the European Health and Digital Executive Agency. Neither the European Union nor the granting authority can be held responsible for them.
\end{acks}

\balance

\bibliographystyle{ACM-Reference-Format}
\bibliography{references}

\end{document}